\documentclass[useAMS,usenatbib]{mn2e}

\usepackage{psfig,epsfig}
\usepackage{mathrsfs}
\usepackage{amssymb}
\usepackage{epsfig}
\usepackage{marvosym}
\usepackage{subfigure}
\usepackage{fancyhdr}
\usepackage{rotating}

\def\degr{\hbox{$^\circ$}}

\def\gsim{\mathrel{\hbox{\rlap{\lower.55ex \hbox {$\sim$}}
                   \kern-.3em \raise.4ex \hbox{$>$}}}}
\def\lsim{\mathrel{\hbox{\rlap{\lower.55ex \hbox {$\sim$}}
                   \kern-.3em \raise.4ex \hbox{$<$}}}}

\def\he{\hbox{He\,{\sc i} $\lambda$5875}}

\def\EGAPS{\hbox{\sl EGAPS\ }}
\def\UVEX{\hbox{\sl UVEX\ }}
\def\IPHAS{\hbox{\sl IPHAS\ }}
\def\RL1{\hbox{{$R_{L_{1}}$}}}

\title[Spectroscopic follow-up of UV-excess objects selected from the UVEX survey]
{Spectroscopic follow-up of UV-excess objects selected from the UVEX survey}
\author[Kars Verbeek et al.]{{Kars Verbeek$^{1}$\thanks{E-mail:k.verbeek@astro.ru.nl}, 
Paul J. Groot$^{1}$,
Simone Scaringi$^{1}$,
Ralf Napiwotzki$^{2}$,}
\newauthor{
Ben Spikings$^{2}$,
Roy H. {\O}stensen$^{3}$,
Janet Drew$^{2}$,
Danny Steeghs$^{4}$,
Jorge Casares$^{5}$,}
\newauthor{
Jesus M. Corral-Santana$^{5,6}$
Romano Corradi$^{5,7}$,
Niall Deacon$^{8}$
Jeremy Drake$^{9}$}
\newauthor{
Boris T. G{\"a}nsicke$^{4}$,
Eduardo Gonz{\'a}lez-Solares$^{10}$,
Robert Greimel$^{11}$,
Ulrich Heber$^{12}$,}
\newauthor{
Mike Irwin$^{10}$,
Christian Knigge$^{13}$
and Gijs Nelemans$^{1}$}\\
$^{1}$Department of Astrophysics, Radboud University Nijmegen,
  P.O. Box 9010, 6500 GL Nijmegen, The Netherlands\\
$^{2}$Centre for Astronomy Research, Science \& Technology Research
  Institute, University of Hertfordshire, Hatfield, AL10 9AB, UK\\
$^{3}$Instituut voor Sterrenkunde, KU Leuven, Celestijnenlaan 200D, B-3001 
  Leuven, Belgium\\
$^{4}$Physics Department, University of Warwick, Coventry, CV4 7AL,
  UK\\
$^{5}$Instituto de Astrof\'{\i}sica de Canarias, Via Lactea, s/n
  E-38205 La Laguna (Tenerife), Spain\\
$^{6}$Departamento de Astrof\'{\i}sica, Universidad de La Laguna, 
  La Laguna E-38205, S/C de Tenerife, Spain\\
$^{7}$Isaac Newton Group of Telescopes, Apartado de Correos 321,
  E-38700 Santa Cruz de La Palma, Canary Islands, Spain\\
$^{8}$Institute for Astronomy, University of Hawaii, 
  2680 Woodlawn Drive, Honolulu, HI 96822, USA\\
$^{9}$Harvard Smithsonian CfA, HCO/SAO, MS 3, 60 Garden Str, US
  Cambridge MA 02138-1516, United States\\
$^{10}$Cambridge Astronomy Survey Unit, Institute of Astronomy, University of
  Cambridge, Madingley Road, Cambridge, CB3 0HA, UK\\
$^{11}$Institut f\"ur Physik, Karl-Franzen Universit\"at Graz,
Universit\"atsplatz 5, 8010 Graz, Austria\\
$^{12}$Dr. Remeis-Sternwarte Bamberg, Universit\"at Erlangen-N\"urnberg,
  Sternwartstrasse 7, D-96049 Bamberg, Germany\\
$^{13}$School of Physics and Astronomy, University of Southampton,
  Southampton, Hampshire, SO17 1BJ, UK
}

\begin{document}

\date{Accepted ...  Received ...; in original form ...}

\pagerange{\pageref{firstpage}--\pageref{lastpage}} \pubyear{2012}

\maketitle

\label{firstpage}

\begin{abstract}
We present the results of the first spectroscopic follow-up of 132 optically blue UV-excess sources 
selected from the UV-excess survey of the Northern Galactic Plane (\UVEX). 
The UV-excess spectra are classified into different populations and grids of model spectra are fit 
to determine spectral types, temperatures, surface gravities and reddening.
From this initial spectroscopic follow-up 95$\%$ of the UV-excess
candidates turn out to be genuine UV-excess sources such as white dwarfs, white dwarf binaries, subdwarfs type O and B, 
emission line stars and QSOs.
The remaining sources are classified as slightly reddened main-sequence stars
with spectral types later than A0V.
The fraction of DA white dwarfs is 47$\%$ with reddening smaller than $E(B-V)$$\leq$0.7 mag.
Relations between the different populations and their \UVEX photometry, Galactic latitude 
and reddening are shown. A larger fraction of \UVEX white dwarfs is found at magnitudes fainter than $g$$>$17 and 
Galactic latitude smaller than $|b|<$4 compared to main-sequence stars, blue horizontal branch stars and subdwarfs.
\end{abstract}


\newpage

\section{Introduction}
Traditionally, surveys searching for faint blue objects have avoided the Galactic Plane because 
of the high dust absorption. Surveys searching for quasars and white dwarfs 
therefore mostly observed at Galactic latitudes larger than $|b|>$30\degr.
Examples of such surveys are
the Palomar Green survey (PG, Green et al., 1986), 
the Kiso survey (Wegner et al., 1987, Limoges et al., 2010),  
the Sloan Digital Sky survey (SDSS, York et al., 2000, Yanni et al., 2009 and Eisenstein et al., 2006)
and the Hamburg Quasar survey (HQS, Hagen et al., 1995, Homeier et al., 1998)
in the northern hemisphere and
the Montreal-Cambridge-Tololo survey (MCT, Lamontagne et al., 2000, Demers 1986), 
the Edinburgh-Cape survey (EC, Kilkenny et al., 1997, Stobie et al., 1997), 
the Homogeneous Bright Quasar survey (Gemmo et al., 1995) and 
the Hamburg-ESO survey (Christlieb et al., 2001, Wisotzki et al., 1996) 
in the southern hemisphere. 
Only the Kitt Peak-Downes survey (KPD, Downes et al., 1986) survey and the Sandage 
Two-colour Galactic Plane survey (Lanning, 1973) observed a bit closer to the Galactic Plane. Some of the brighest \UVEX UV-excess 
sources are Lanning sources (e.g. UVEXJ0328+5035 and UVEXJ0528+2716 in Table\ \ref{tab:spectra} are in Lanning, 1973 and Lanning et al., 2004 respectively).
The lowest Galactic latitudes $|b|<5\degr$ are still relatively unexplored (see e.g. Fig.2 of Napiwotzki et al., 2003).
In order to determine key population characteristics of Galactic sources, such as their
scaleheight or space density, it is crucial to study the low Galactic latitude environment.
The space density of stellar remnants, such as white dwarfs, Cataclysmic Variables and AM CVn stars, 
is currently poorly constrained while there must be $\sim10^{5}$ of them in the Milky Way 
(see Fig.\,1 of Groot et al., 2009, McCook et al., 1999, L\'epine et al., 2011 and Nelemans et al., 2001).\\

One of the main goals of the European Galactic Plane Surveys (\EGAPS) is to obtain a homogeneous sample 
of evolved objects in our Milky Way with well-known selection limits.
The UV-excess survey of the Northern Galactic Plane (\UVEX, Groot et al., 2009) 
images a 10$\times$185 degrees wide band (--5\degr$<$ $b$ $<$+5\degr) centred on the Galactic equator
in the $U,g,r$ and $\he$ bands down to $\sim 21^{st}-22^{nd}$ magnitude using the Wide Field Camera mounted 
on the Isaac Newton Telescope on La Palma. From the first 211 square degrees of \UVEX data, 
a catalogue of 2\,170 optically blue UV-excess candidates was selected in Verbeek et al. (2012; hereafter V12). 
These UV-excess sources were selected from the $(U-g)$ versus $(g-r)$ colour-colour diagram 
and $g$ versus $(U-g)$ and $g$ versus $(g-r)$ colour-magnitude diagrams 
by an automated field-to-field selection algorithm.
This automated selection algorithm and the properties of the selected UV-excess catalogue are described in V12.
Less than $\sim$1$\%$ of the selected UV-excess sources are currently known in the literature.\\

Here we report our spectroscopic follow-up for 132 objects (6$\%$) in the UV-excess catalogue of V12.
This early reconnaissance is important for the design of future 
colour-selection methods for various populations, comparable to the selection techniques for e.g. the SDSS, which generally
do not have to deal with the added complication of reddening (G\"ansicke et al., 2009, Girven et al., 2011).
In Sect.\ \ref{sec:followup} the spectroscopy of the selected sample is described,
and in Sect.\ \ref{sec:spectra} the spectra are presented and classified.
The spectra are fitted to grids of model spectra in order to determine the characteristics of UV-excess spectra classified as white dwarfs, 
subdwarfs, main-sequence stars and blue horizontal branch stars.
Finally in Sect.\ \ref{sec:discussion} we summarise the conclusions of the UV-excess catalogue and the spectroscopic follow-up.
The UV-excess spectra are shown in Figs.\ \ref{fig:spectra1} to\ \ref{fig:spectra11} and their features are listed 
in Table\ \ref{tab:spectra} in Appendix\ \ref{app:appendix}. All spectra and the table can also be 
obtained from the \UVEX website\footnote{http://www.uvexsurvey.org}.\\

\begin{figure*}
\centerline{\epsfig{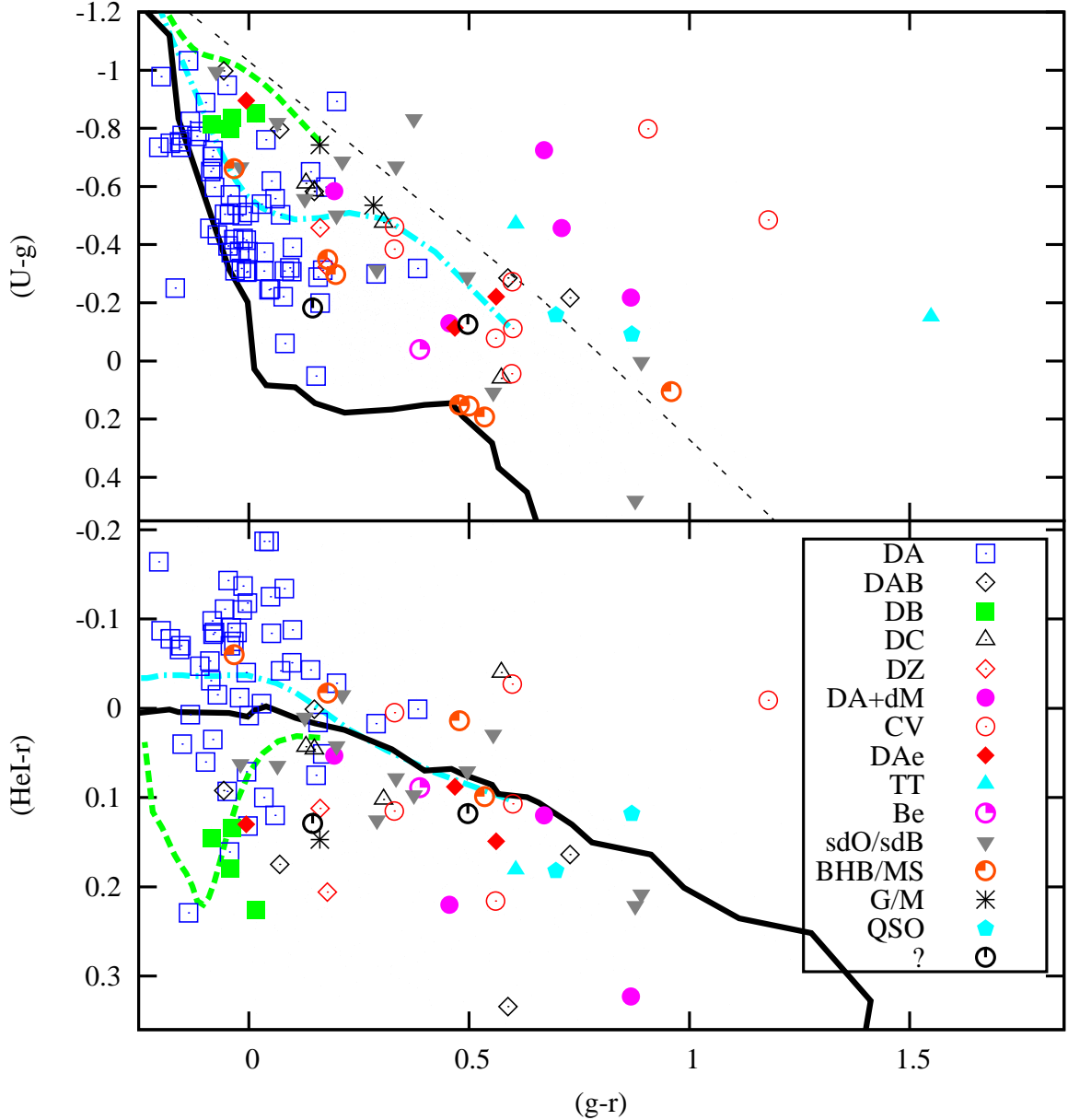}}
\caption{The UVEX colour-colour diagrams with the classified UV-excess candidates. 
The lines are the simulated colours of unreddened main-sequence stars (solid black) and the O5V-reddening line (dashed black) of V12. 
The cyan and green dashed lines are respectively the simulated colours of unreddened Koester DA and DB white dwarfs. 
The different symbols indicate the classification: White Dwarf (DA/DB/DAB/DC/DZ/DAe), White Dwarf+Red Dwarf binary (DA+dM), 
Cataclysmic Variable (CV), T Tauri star (TT), Be star (Be), subdwarf star (sdO/sdB), main-sequence star or blue horizontal branch star (MS/BHB),
G2V star and M-giant (G/M), Quasi Stellar Object (QSO) and unknown (?). The sources classified as ``noisy'' in Sect.\ \ref{sec:mainseqsubdwarf} are not shown here.
There is one more H$\alpha$ emitter classified as T Tauri star at $(g-r)$=1.55, $(HeI-r)$=0.6 and one M-giant at $(g-r)$=0.28, $(HeI-r)$=--1.9
not shown in the $(HeI-r)$ vs. $(g-r)$ colour-colour diagram.
\label{fig:classCCD}}
\end{figure*}

\begin{figure*}
\centerline{\epsfig{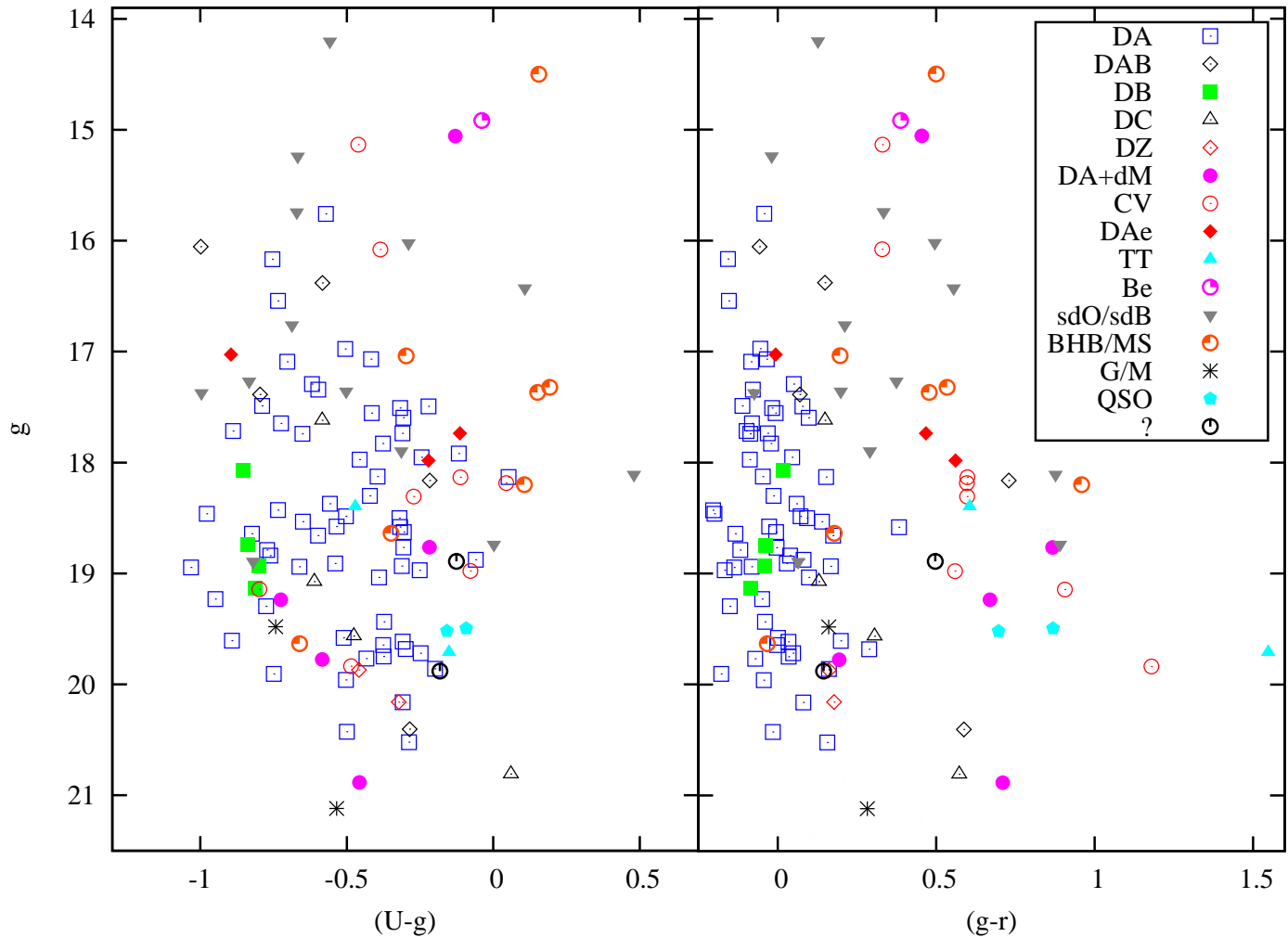}}
\caption{The UVEX colour-magnitude diagrams with the classified UV-excess candidates.
\label{fig:classCMD}}
\end{figure*}

\begin{figure*}
\centerline{\epsfig{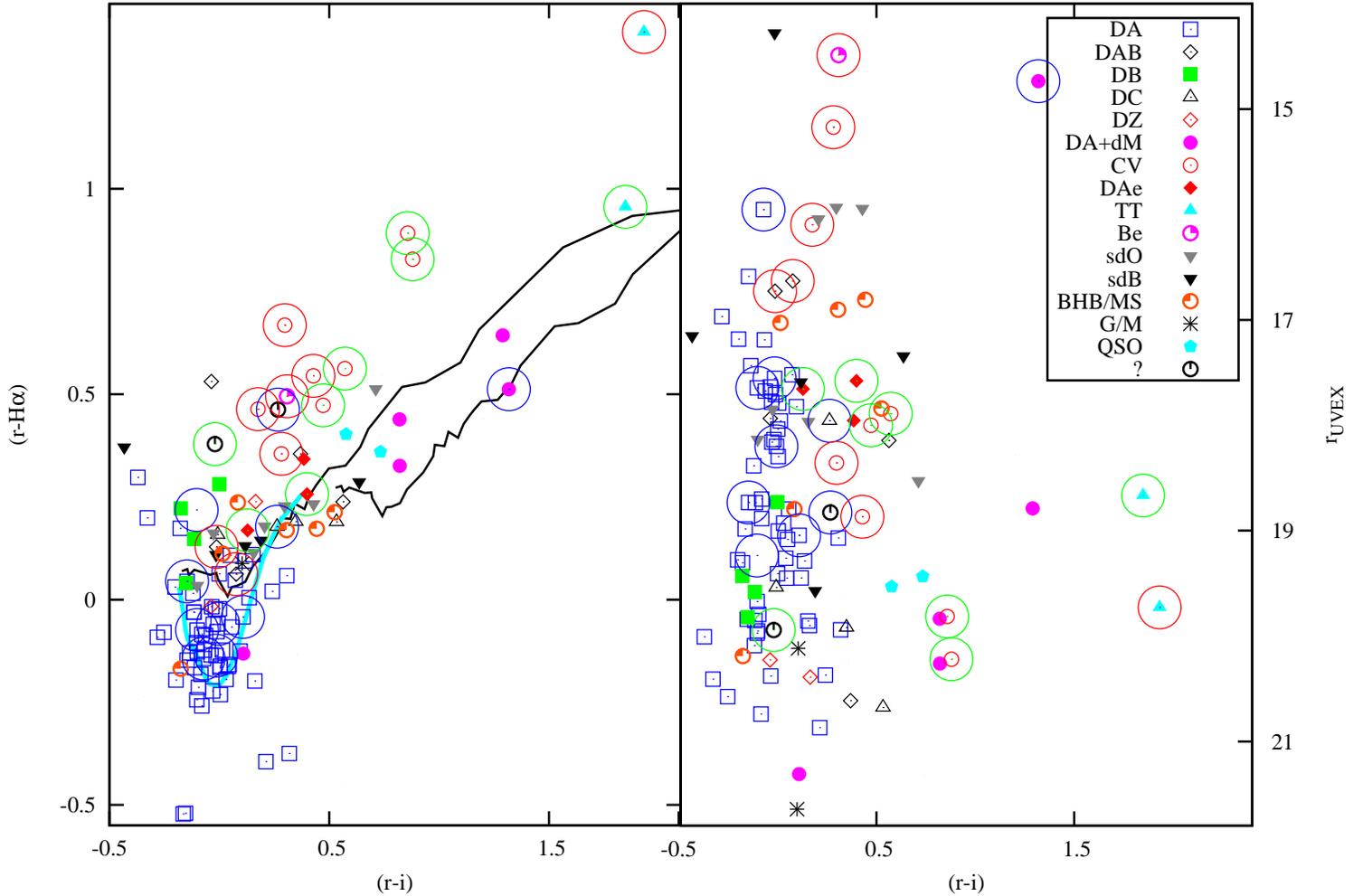}}
\caption{IPHAS colour-colour and colour-magnitude diagrams with the classified UV-excess candidates 
that have a match in IPHAS. There is one extra sources classified as M-giant at $(r-i)$=0.1, $(r-H\alpha)$=2.25 
in the $(r-H\alpha)$ vs. $(r-i)$ colour-colour diagram. The lines are the synthetic colours of main-sequence stars (black) 
with reddening $E(B-V)$=0 and $E(B-V)$=1 and unreddened Koester DA white dwarfs (cyan).
Sources that are in the Deacon IPHAS-POSSI PM catalogue are encircled blue, 
sources that are in the Witham H$\alpha$ emission line catalogue are encircled red and
sources that show H$\alpha$ emission lines in their spectra but are not in the Witham catalogue are encircled green.
The Witham catalogue covers the magnitude range 13$<r<$19.5 and the Deacon catalogue 
covers the magnitude range 13.5$<r<$19. The sources classified as ``noisy'' in Sect.\ \ref{sec:mainseqsubdwarf} are not shown here, 
one of them has a match in the IPHAS-POSSI PM catalogue.
\label{fig:IPHASCCD}}
\end{figure*}

\section{Spectroscopic follow-up of UV-excess candidates}
\label{sec:followup}
Spectroscopic follow-up was obtained by three different telescopes for a total of 132 UV-excess candidates during a number of observing runs.
For 100 UV-excess candidates spectroscopic observations were obtained, during two runs in September 2009 
and December 2010, with the Intermediate dispersion Spectrograph and Imaging System (ISIS) mounted at the 4.2m 
William Herschel Telescope (WHT) at Roque de los Muchachos Observatory, on the island of La Palma.
The blue and red arms of the spectrograph were used in combination with the standard 5300 dichroic and no order sorting filter. 
The gratings $R300B$ in the blue arm and $R316R$ in the red arm were used giving a dispersion of 0.86 \AA/pix and 0.93 \AA/pix, respectively.
The central wavelengths of the blue and red arms were $\lambda_{c}$=4\,700 \AA\, and $\lambda_{c}$=6\,650 \AA, respectively. 
The slit width (1.2-1.5 arcsec) was matched with the seeing during the observations: typically 
20-30 percent larger than the seeing. The binning was 2$\times$2 and the read-out speed slow.
We used integration times from 300 seconds for the brightest objects at $g$$\sim$15 
to 1500 seconds for the fainter sources at $g$$\sim$20. This gives signal-to-noise ratio SNR$\geq$20, 
which is required for spectroscopic identification of the UV-excess sources.
The goal was to obtain a sample of spectra, distributed equally over $g$ magnitude and $(g-r)$ colours in the magnitude range $13<g<20$, 
covering the entire $g$ vs. $(g-r)$ colour-magnitude diagram. Due to the weather and the location of the Galactic Plane during the observations 
the sample is biased in magnitude and right ascension. For statistics it is important to be aware of this bias.\\

All the WHT/ISIS spectra were reduced using 
IRAF\footnote{Image Reduction and Analysis Facility (IRAF) 
is distributed by the National Optical Astronomy Observatory, 
which is operated by the Association of Universities for Research in Astronomy (AURA) 
under cooperative agreement with the National Science Foundation.}. 
Bias and flat field corrections, trimming and extraction of the spectra were done in the standard way. 
The spectra were wavelength calibrated using CuNe+CuAr calibration arcs. 
Standard stars BD+28$^\circ$4211, BD+25$^\circ$4655, G191--B2B, Feige 34 and Feige 110
were used for the flux calibration of the spectra. The spectra were not corrected for telluric
absorption. Effects of cosmic rays not removed by IRAF were corrected by hand by interpolating these pixels to the 
average flux of the neighbouring pixels. 
The reduced WHT/ISIS spectra cover the wavelength range $\lambda$=3\,700 \AA\, to $\lambda$=8\,100 \AA, 
with a dichroic gap from $\lambda$=5\,200--5\,600 \AA. 
Two additional WHT/ISIS spectra with similar characteristics were obtained during a run in October 2008 
during follow-up of IPHAS-POSS high proper motion candidates (Deacon et al., 2009).\\

Furthermore, twenty-six Hectospec (Fabricant et al., 2004) spectra are available for the UV-excess candidates. These 26 spectra were obtained 
with the MMT+Hectospec combination during \IPHAS follow-up observations between 2004 and 2007, described in Sect\,2.1 of Vink et al. (2008).
Hectospec is a multi-object spectrograph, fed by 300 robotically-positioned optical fibers, 
attached to the 6.5m MMT telescope on Mount Hopkins, Arizona, USA.
The spectra cover the wavelength range $\lambda$=4\,000--8\,500 \AA\, and have a dispersion of $\sim$6 \AA/pix.
Of the 26 Hectospec spectra 5 spectra are flux calibrated. 
The extracted Hectospec were corrected for incomplete sky subtraction (Vink et al., 2008) 
and background sky spectra were checked in order to confirm the emission lines.
Some of the Hectospec spectra shown in Appendix\ \ref{app:appendix} still show emission features
at the wavelengths of the Balmer lines due to bad sky subtraction in fields with diffuse emission (Fabricant et al., 2005).\\

Additionally, the FAST spectrograph (Fabricant et al., 1998), mounted on the 60-inch Tillinghast telescope, 
located at the Fred Lawrence Whipple Observatory (FLWO) on Mount Hopkins, Arizona, 
obtained spectra for candidates in the \IPHAS H$\alpha$ emission line list (Witham et al., 2008) and 
candidates in the catalogues of V12. There are 4 FAST spectra for our UV-excess candidates obtained between 2009 and 2012. 
The FAST spectra cover the wavelength range $\lambda$=3\,800--7\,400 \AA\, with a dispersion of $\sim$3 \AA/pix.\\

\newpage

\section{The classification of the spectroscopic observations}
\label{sec:spectra}
The results of the spectroscopic observations are presented in Table\ \ref{tab:spectra} in Appendix\ \ref{app:appendix}, ordered by RA.
The spectra of all UV-excess sources are shown in Figs.\ \ref{fig:spectra1} to\ \ref{fig:spectra11} in Appendix\ \ref{app:appendix}. 
An overview of the classification is summarized in Table\ \ref{tab:spectraltypes} and the classified sources are plotted in 
the colour-colour and colour-magnitude diagrams of Figs.\ \ref{fig:classCCD} to \ref{fig:IPHASCCD}.
The INT/WFC Photometric H$\alpha$ Survey of the Northern Galactic Plane (\IPHAS, Drew et al., 2005)
imaged the same survey area as \UVEX in the $r$, $i$ and $H\alpha$ filters, the \IPHAS IDR data 
(Gonz\'{a}lez-Solares et al., 2008) are used in Fig.\ \ref{fig:IPHASCCD}.
In the \IPHAS colour-colour and colour-magnitude diagram sources with a match in Witham H$\alpha$ emission line catalogue (Witham et al., 2008)
or IPHAS-POSS proper motion catalogue (Deacon et al., 2009) are encircled red and blue respectively.
Note that a global photometric calibration is not applied to the \UVEX data 
yet, so the magnitudes and colours of the UV-excess sources might show a small scatter 
(similar to the early \IPHAS data, e.g. Drew et al., 2005).
Additionally, there is a likely systematic shift in $(U-g)$ of 0.2-0.3 magnitudes for all sources. 
This $U$-band shift in the INT/WFC data was already reported in Greiss et al. (2012). 
Both effects do not influence the result of the selection method and the content of the UV-excess catalogue because 
the selection was done relative to the reddened main-sequence population (see V12 for details), 
however the shift does apply to the $(U-g)$ colours given in Table\ \ref{tab:spectra}.\\

\begin{table}
\caption[]{Classification of UV-excess spectra. \label{tab:spectraltypes} }
\centering
{\small
\begin{tabular}{ | l | r | r | }
    \hline
Spec.Type    &  Number &  Fraction ($\%$)  \\ \hline
DA	     &	 62   &  46.6  \\
DB	     &    4    &  3.0	\\
DAB/DBA      &    5    &  3.8	\\
DC	     &	  4    &  3.0   \\
DZ	     &	  2    &  1.5   \\
DA+dM	     &	  5    &  3.8   \\
DAe	     &    3    &  2.3	\\
CV	     &	  8    &  6.0   \\
T Tauri      &    2    &  1.5	\\
Be	     &	  1    &  0.8   \\
sdO	     &	  7    &  5.3   \\
sdB	     &	  6    &  4.5   \\
BHB/MS/F     &    4    &  3.0	\\
BHB/MS/B     &    3    &  2.3	\\
G/M	     &    2    &  1.5	\\
QSO	     &	  2    &  1.5   \\
Unknown      &    2    &  1.5	\\
Noisy        &    10   &  7.6   \\
    \hline
\end{tabular} \\ 
}
\end{table}


\begin{figure}
\centerline{\epsfig{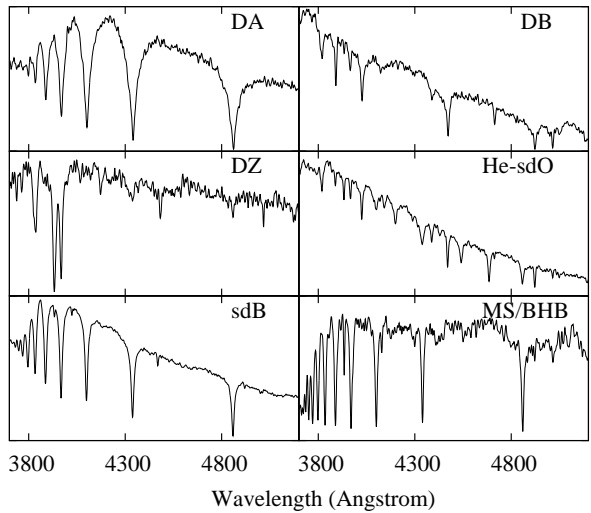}}
\caption{Example of 6 UV-excess spectra: DA white dwarf (UVEXJ0113+5819), DB white dwarf (UVEXJ0002+6236),
DZ white dwarf (UVEXJ0418+4417), He-sdO (UVEXJ0221+5648), sdB (UVEXJ0202+5643) and MS/BHB star (UVEXJ0228+5846).
\label{fig:examplespectra}}
\end{figure}

\subsection{The UV-excess spectra classified as white dwarfs}
\label{sec:whitedwarf}
A first classification of the UV-excess spectra is done by comparing them with model spectra by eye.
White dwarfs and emission-line star spectra are separated from the subdwarf type O and B, main-sequence and blue horizontal branch star spectra
(see Sect.\ \ref{sec:mainseqsubdwarf}).
Hydrogen atmosphere (DA) white dwarfs are easily recognizable by their broad Balmer lines. We use the 
classification criteria of Sion et al. (1983) and the atlas of Wesemael et al. (1993) to classify the other different types
of white dwarfs by eye. In total we classify 85 spectra as white dwarfs.
Sixty-two show only Balmer lines (DA), there are 4 white dwarfs
showing only HeI lines (DB) and 5 white dwarfs showing both Balmer and HeI lines (DBA/DAB). 
Four white dwarfs have a continuum spectrum with no lines (DC) and 2 white dwarfs show 
a spectrum with strong calcium lines only but no, or only little, hydrogen and helium lines (DZ/DZA).
Five objects are DA+dM composite objects, showing a DA white dwarf in the blue part of the spectrum 
and an M-dwarf in the red part of the spectrum (Lanning, 1982).
Furthermore, there are 3 sources showing a DA white dwarf spectrum, with emission lines at the centre of their Balmer absorption lines. 
These DAe sources are discussed in Sect.\ \ref{sec:emissionline}. 
Example spectra of a DA white dwarf, DB white dwarf and a DZ white dwarf are shown in Fig.\ \ref{fig:examplespectra}.\\

\begin{figure}
\centerline{\epsfig{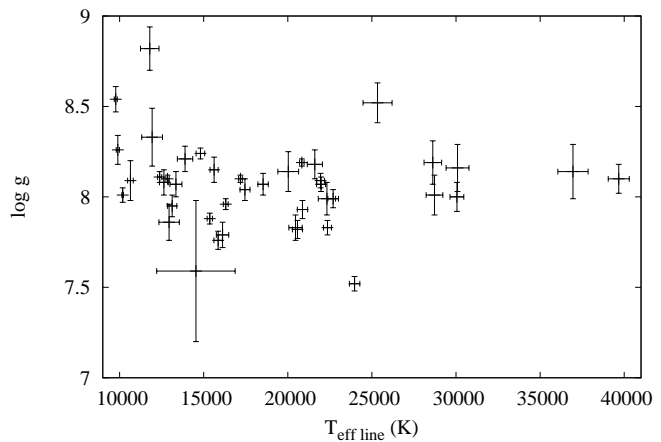}}
\caption{Temperature-gravity diagram of the UV-excess hydrogen atmosphere white dwarfs, determined through line profile fitting.
\label{fig:gravitytemperaturewds}}
\end{figure}

\begin{figure}
\centerline{\epsfig{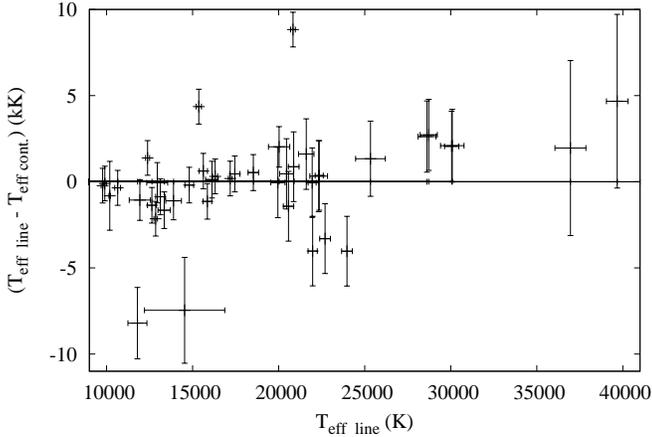}}
\caption{Comparison of the hydrogen white dwarf temperatures found by the two independent fitting methods:
Method 1 fits the absorption lines of the normalized spectra and method 2 
fits the continuum of the spectra including reddening.
\label{fig:comparetempfitwds}}
\end{figure}

\begin{table}
\caption[]{Result of the two fitting methods for WHT/ISIS UV-excess spectra classified as white dwarfs.
The first method fits $T_{\rm eff}$ and $log\,g$, the second method fits $T_{\rm eff}$ and $E(B-V)$.
 \label{tab:whitedwarfs}}
{\scriptsize
\begin{tabular}{ | l | c | c | c | }
    \hline
ID	&	UVEX name   		&  $T_{\rm eff}$(K)/$log\,g$          &  $T_{\rm eff}$(K)/$E(B-V)$   \\ \hline
1      &    011311.87+581902.3      &					 &   10\,000/0.0  \\
2      &    011754.90+581815.4      &  11\,933$\pm$619 /8.33$\pm$0.16	  &   13\,000/0.0  \\
3      &    012015.68+584318.2      &  36\,955$\pm$899 /8.14$\pm$0.15	  &   35\,000/0.2  \\
4      &    012219.81+611229.9      &  12\,801$\pm$132 /8.1 $\pm$0.03	  &   11\,000/0.2  \\
5      &    012359.82+672223.1      &					 &   57\,000/0.7  \\
6      &    020201.82+564744.8      &					 &   65\,000/0.3  \\
7      &    022135.47+564436.6      &  10\,183$\pm$46 /8.01$\pm$0.04	 &   22\,000/0.2  \\
8      &    022151.40+563815.7      &  28\,621$\pm$522 /8.19$\pm$0.12	  &   26\,000/0.1  \\
9      &    022510.84+580156.6      &  13\,124$\pm$274 /7.95$\pm$0.06	  &   14\,000/0.0  \\
10     &    022615.13+581710.2      &  9\,772 $\pm$90 /8.54$\pm$0.07	 &   10\,000/0.0  \\
11     &    023044.92+563622.6      &  30\,091$\pm$682 /8.16$\pm$0.13	  &   28\,000/0.1  \\
12     &    032737.64+530231.1      &  18\,531$\pm$306 /8.07$\pm$0.06	  &   18\,000/0.1  \\
13     &    032807.05+525737.2      &  23\,971$\pm$310 /7.84$\pm$0.04	  &   28\,000/0.3  \\
14     &    032908.01+524400.6      &  22\,358$\pm$235 /7.83$\pm$0.04	  &   22\,000/0.2  \\
15     &    032910.60+524426.3      &  12\,852$\pm$102 /8.10$\pm$0.02	  &   15\,000/0.0  \\
16     &    033118.06+530351.3      &  17\,185$\pm$112 /8.10$\pm$0.02	  &   17\,000/0.0  \\
17     &    041045.70+461137.1      &  14\,537$\pm$2333 /7.59$\pm$0.39    &   22\,000/0.0  \\
18     &    041053.99+450706.5      &  16\,122$\pm$370 /7.79$\pm$0.07	  &   16\,000/0.1  \\
19     &    041359.37+455151.2      &  12\,381$\pm$121 /8.11$\pm$0.03	  &   11\,000/0.0  \\
20     &    041733.05+452524.4      &  22\,697$\pm$314 /7.99$\pm$0.05	  &   26\,000/0.2  \\
21     &    041902.55+434307.1      &  17\,454$\pm$293 /8.04$\pm$0.06	  &   17\,000/0.1  \\
22     &    042110.67+440945.6      &  20\,574$\pm$300 /7.82$\pm$0.05	  &   22\,000/0.4  \\
23     &    052825.82+320859.5      &  9\,894 $\pm$94 /8.26$\pm$0.08	 &   10\,000/0.0  \\
24     &    052847.75+322330.3      &  12\,942$\pm$607 /7.86$\pm$0.10	  &   13\,000/0.0  \\
25     &    190812.07+164029.2      &					 &   17\,000/0.0  \\
26     &    202249.99+412423.1      &  15\,620$\pm$240 /8.15$\pm$0.07	  &   15\,000/0.4  \\
27     &    202255.55+412504.9      &  16\,309$\pm$146 /7.96$\pm$0.03	  &   16\,000/0.0  \\
28     &    202350.92+423826.0      &  25\,335$\pm$864 /8.52$\pm$0.11	  &   24\,000/0.1  \\
29     &    202439.91+400630.7      &					 &   24\,000/0.0  \\
30     &    202457.34+410804.1      &					 &   22\,000/0.1  \\
31     &    202501.86+411626.0      &  14\,805$\pm$252 /8.24$\pm$0.03	  &   15\,000/0.3  \\
32     &    202557.21+400949.2      &					 &   30\,000/0.1  \\
33     &    202800.47+405620.0      &					 &   20\,000/0.0  \\
34     &    203739.63+413216.3      &					 &   17\,000/0.3  \\
35     &    205037.81+424618.9      &					 &   17\,000/0.1  \\
36     &    205148.13+442408.8      &					 &   26\,000/0.0  \\
37     &    210037.77+501029.0      &  20\,873$\pm$308 /7.93$\pm$0.05	  &   20\,000/0.0  \\
38     &    210248.44+475058.9      &  13\,348$\pm$359 /8.07$\pm$0.07	  &   15\,000/0.0  \\
39     &    211718.18+550638.7      &					 &   22\,000/0.0  \\
40     &    212409.05+555521.4      &					 &   12\,000/0.0  \\
41     &    212852.14+542048.4      &  13\,891$\pm$457 /8.21$\pm$0.07	  &   15\,000/0.0  \\
42     &    222940.17+610700.7      &  22\,325$\pm$515 /7.99$\pm$0.09	  &   22\,000/0.0  \\
43     &    223634.77+591907.8      &  28\,721$\pm$497 /8.01$\pm$0.11	  &   26\,000/0.1  \\
44     &    223811.54+603759.9      &  12\,633$\pm$261 /8.08$\pm$0.07	  &   14\,000/0.0  \\
45     &    224010.23+555950.6      &  19\,957$\pm$414 /3.03$\pm$0.07	  &   20\,000/0.2  \\
46     &    224610.82+611450.3      &  20\,026$\pm$617 /8.14$\pm$0.11	  &   18\,000/0.0  \\
    \hline
\end{tabular} \\ 
}
\end{table}

\begin{figure*}
\centerline{\epsfig{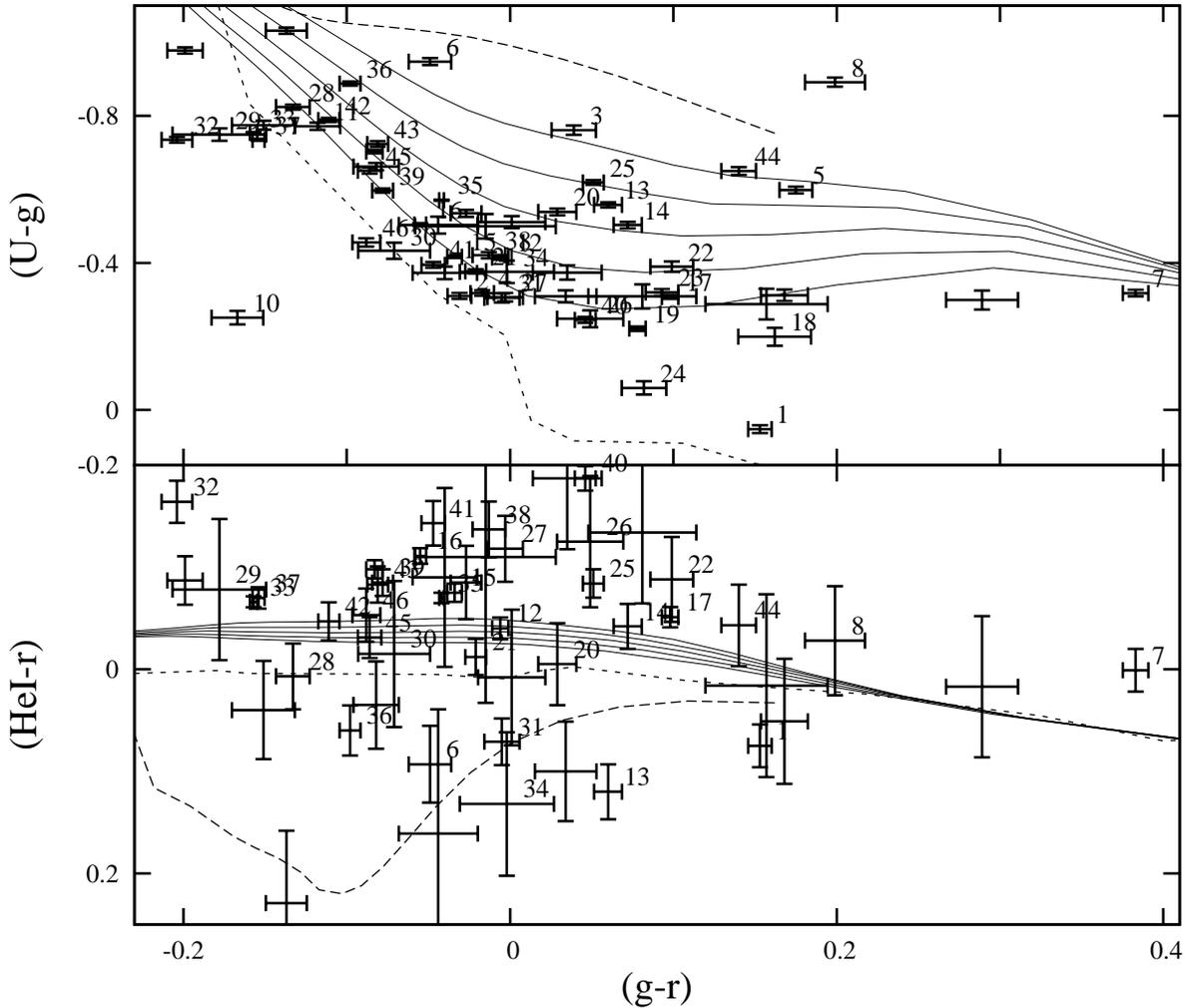}}
\caption{The UVEX colour-colour diagrams with all UV-excess sources classified as hydrogen white dwarfs, plotted with their photometric error bars. 
The lines are the simulated colours of unreddened main-sequence stars (dotted) and unreddened Koester DB white dwarfs (dashed) of V12
and the numbers are the white dwarf object IDs of Table\ \ref{tab:whitedwarfs}.
The five solid lines are the simulated colours of unreddened Koester DA white dwarfs with $log\,g$=7.0,7.5,8.0,8.5,9.0
(in both colour-colour diagrams the line with $log\,g$=9.0 is the most upper line).
Data points and simulated colours show a likely systematic shift in $(U-g)$ colours, as explained in Greiss et al., 2012.
Note that a global photometric calibration is not applied to the \UVEX data yet, due to a scatter in the $HeI$-band photometry  
the white dwarfs do not overlap the simulated white dwarfs colours in the $(HeI-r)$ vs. $(g-r)$ colour-colour diagram. 
The spectra of the DA white dwarfs above or below the simulated white dwarf colours in the $(HeI-r)$ vs. $(g-r)$ colour-colour diagram 
do not show HeI emission or absorption.
\label{fig:whitedwarfCCDs}}
\end{figure*}

\begin{figure}
\centerline{\epsfig{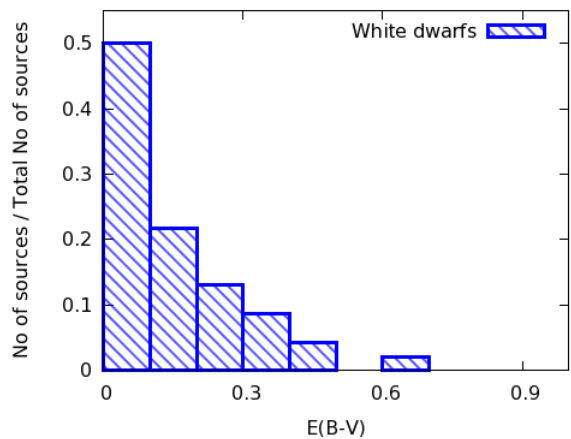}}
\caption{Distribution of $E(B-V)$ fit to the WHT/ISIS spectra of the sources classified as hydrogen atmosphere white dwarfs.
The number of sources per bin is normalized by the total number of white dwarfs.
\label{fig:reddeningwds}}
\end{figure}

White dwarf model spectra are fitted to determine the effective 
temperature ($T_{\rm eff}$) and surface gravity ($log\,g$) of the DA white dwarfs by two independent methods. 
The first method, described in Napiwotzki (1997) and Napiwotzki et al. (1999), normalizes the continuum of the white dwarf spectra and 
then fits the absorption lines using an interpolated grid of model spectra with different $T_{\rm eff}$ and $log\,g$
at $\Delta T_{\rm eff}$=500--2\,000 K and $\Delta log\,g$=0.2/0.5 intervals.
The second method fits a grid of reddened white dwarfs model spectra (Koester et al., 2001) with $log\,g$=8.0 to 
the spectra in the range 0.0$\leq$$E(B-V)$$\leq$1.0 at $\Delta E(B-V)$=0.1 intervals, using the reddening laws of Cardelli, Clayton \& Mathis (1989).
The second method fits the effective temperature and reddening of the white dwarfs with an accuracy of 
$T_{\rm eff}$$\sim$1\,000K and $E(B-V)$$\sim$0.1 (see Sect.\ \ref{sec:discussion}).
The Hectospec white dwarf spectra are not fitted since they are not flux calibrated.
The white dwarf fitting results of the two methods are listed in Table\ \ref{tab:whitedwarfs}.
The result of the first white dwarf fitting method is shown in the temperature-gravity diagram of Fig.\ \ref{fig:gravitytemperaturewds},
The difference between the two methods is shown in Fig.\ \ref{fig:comparetempfitwds}, where typically the results agree within the errors.
Four systems do not have consistent fits, for these spectra the continuum fit of the second method seems by eye to be the most suitable.
For fitting of normalised profiles there is a strong degeneracy between
a ``hot'' and ``cold'' solution, due to a similar equivalent width of the Balmer
lines, here the second method is more robust regarding the temperature since the slope is taken into account.
The object IDs from Table\ \ref{tab:whitedwarfs} are overplotted in the colour-colour diagrams of Fig.\ \ref{fig:whitedwarfCCDs}.
The reddenings found by the second fitting method are shown in the histogram of Fig.\ \ref{fig:reddeningwds}.
Hydrogen white dwarfs typically show a reddening of 0.0$\leq$$E(B-V)$$\leq$0.1
with a reddening up to $E(B-V)$$\leq$0.7 for the hotter white dwarfs.
As expected given their intrinsic luminosities, we see hot white dwarfs out to
larger distances compared to cool white dwarfs, thus hot white dwarfs typically suffers from slightly larger reddening compared to 
cooler white dwarfs. Cool white dwarfs can only have little reddening since they are an intrinsically faint local population.\\

\begin{table}
\caption[]{Result of the fitting for UV-excess spectra classified as subdwarfs, main-sequence stars (MS) and blue horizontal branch stars (BHB). 
The third column shows for the He-sdO, sdO, sdB and B-type BHB/MS sources the best fit: $T_{\rm eff}$ in kK, $log\,g$ and $log\,(n(He)/n(H))$. 
For the MS/BHB sources without He lines the third column shows $T_{\rm eff}$ in kK and $log\,g$.\label{tab:mainseqsubdwarf}}
{\small
\begin{tabular}{ | l | c | l | }
    \hline
UVEX name    		 & 	$g$        &  Classification/fitting   \\ \hline
UVEXJ000016.27+603246.3  &     16.761	   & sdB (34.4/6.0/--2.8) \\
UVEXJ001032.27+625050.0  &     16.427      & He-sdO \\
UVEXJ020201.85+564342.3  &     15.238	   & sdB (27.5/5.5/--2.8) \\
UVEXJ022113.52+564810.7  &     17.267	   & He-sdO (44.0/5.5/1.4) \\
UVEXJ022241.76+562702.2  &     17.895	   & sdB/sdO \\
UVEXJ022815.18+584640.8  &     18.200	   & MS/BHB (14.2/3.8) \\
UVEXJ031943.45+512309.0  &     17.368	   & MS/BHB (15.8/4.3) \\
UVEXJ032855.25+503529.8  &     14.202	   & sdB (28.5/5.5/--2.5)  \\
UVEXJ041745.78+454049.8  &     17.322	   & MS/BHB (13.4/3.9) \\
UVEXJ042125.70+465115.4  &     18.107	   & sdB+F composite \\
UVEXJ052835.30+271650.0  &     14.499	   & BHB/MS (17.3/3.7/--1.2) \\
UVEXJ193809.18+305401.5  &     17.038	   & BHB/MS (18.9/4.5/--1.5)  \\
UVEXJ193813.83+313708.1  &     17.373	   & sdO (50.4/5.7/--1.47) \\
UVEXJ193847.06+312024.2  &     19.481	   & G2V, $E(B-V)$=0.3 \\
UVEXJ193951.69+302600.7  &     17.358	   & sdB (27.8/4.6/--1.5) \\
UVEXJ194028.01+322039.8  &     18.638	   & BHB/MS (16.0/3.1) \\
UVEXJ194135.69+321222.0  &     18.892	   & sdB (31.0/6.0/--1.9) \\
UVEXJ204210.22+443928.7  &     18.738	   & sdB/sdO \\
UVEXJ204957.75+401637.9  &     15.741	   & He-sdO (45.9/6.1/1.4) \\
UVEXJ205039.07+373958.4  &     21.119	   & M5III, $E(B-V)$=0.0 \\
UVEXJ223941.98+585729.1  &     16.020	   & He-sdO (47.7/5.0/--0.7) \\
UVEXJ224521.39+551705.3  &     19.632	   & MS/BHB (14.9/4.7) \\
    \hline
\end{tabular} \\ 
}
\end{table}


\subsection{The UV-excess spectra classified as hot subdwarfs, MS stars and BHB stars}
\label{sec:mainseqsubdwarf}
For the classification of hot subdwarfs, main-sequence stars (MS) and blue horizontal branch stars (BHB) stars 
we follow the classification scheme as outlined in Fig.1 of Moehler et al. (1990). Most sources have hydrogen and helium absorption lines 
clearly stronger than the hydrogen and helium absorption lines of their best fit Pickles template spectrum, we classify these sources as hot subdwarfs (sdB/sdO). 
Sources with clear HeII lines are labeled sdO candidates and the sources without HeII lines are labeled sdB candidates.
Grids of model spectra are fitted to the spectra of the sources classified as hot subdwarfs, MS stars and BHB stars as described in {\O}stensen et al. (2011).
The results are listed in Table\ \ref{tab:mainseqsubdwarf} and the spectra of the sources classified as subdwarfs, MS stars and BHB 
stars are shown in Figs.\ \ref{fig:spectra6} and \ref{fig:spectra7}.\\

From the fitting we classify 4 sources as He-sdO stars, 1 sdO star, 5 sdB stars, 1 sdB+F star.
Two more sources are probably sdO/sdB stars but they have no accurate fitting result.
The sdB+F star (UVEXJ0421+4651) has 2MASS (Cutri et al., 2003) photometry $J$=15.7, $H$=15.3, $K$=15.3, so the F/G star dominates in the IR completely.
An example of spectra classified as He-sdO and sdB stars are shown in Fig.\ \ref{fig:examplespectra}.
We classify 7 sources as MS/BHB stars: 4 with $T_{\rm eff}$$<$16kK and 3 B-type MS/BHB stars with $T_{\rm eff}$$>$16kK.
An example of a spectrum classified as MS/BHB star is shown in Fig.\ \ref{fig:examplespectra}.
Due to the signal-to-noise rate (SNR) and resolution of the spectra it is not possible to 
distinguish BHB stars from MS stars. 
Higher SNR and resolution spectra are necessary for a more reliable classification.
Additionally, there is 1 G-type star and 1 M-giant in the spectroscopic sample.
Ten spectra with clear hydrogen absorption lines have no fitting result. 
These sources are not hydrogen atmosphere white dwarfs since the Balmer lines are too narrow, they are probably subdwarfs, MS stars or BHB stars
except for the source UVEXJ2036+3929, which might be a white dwarf.
Since it is not possible to classify these spectra in detail and since most have low SNR they are labeled ``Noisy''. 
These spectra are shown in Fig.\ \ref{fig:spectra11}.\\

Template spectra of main-sequence stars and giants from the library of Pickles et al. (1998) are fit to all main-sequence and blue horizontal branch spectra,
allowing for interstellar reddening in the range 0.0$\leq$$E(B-V)$$\leq$1.0 at $\Delta E(B-V)$=0.1 intervals.
The accuracy of this fitting is discussed in Sect.\ \ref{sec:discussion}.
The fitting of reddened main-sequence stars including their continuum suffers from a well known degeneracy between 
reddened early type stars and unreddened (or less-reddened) late type stars.
We use the characteristic lines of different spectral types (Morgan, Keenan \& Kellman 1943)
and the equivalent width of the CaII K line at $\lambda$=3\,934 \AA\, to confirm the results 
of the fitting method and to break degeneracies where necessary.
The G-type star is a G2V star with reddening $E(B-V)$=0.3 and the M-giant has spectral type M5III with reddening $E(B-V)$=0.0.
We would not expect these G2V and MIII stars in the UV-excess catalogue, they are probably selected due to 
the intrinsic \UVEX photometry scatter. (see Sect.\ \ref{sec:discussion}).\\

\subsection{The UV-excess spectra classified as emission line stars}
\label{sec:emissionline}
Among the 132 UV-excess spectra there are 11 clear H$\alpha$ emission line objects, 
shown in in Figs.\ \ref{fig:spectra8} to\ \ref{fig:spectra9}:
8 Cataclysmic Variables, 2 T Tauri stars and 1 Be star.
We classify the Classical T Tauri from the hydrogen Balmer lines in emission on top of a M-dwarf atmosphere in combination with a U-band excess
and CaII in emission (Corradi et al., 2010, Barentsen et al., 2011). The H$\alpha$ emission lines of the T Tauri candidates have a width of $FWHM$$\sim$5 \AA.
The Be star is classified from the combination of a B-type continuum with Balmer absorption lines and a clear H$\alpha$ emission line.
Additionally, three sources show a hydrogen white dwarf spectrum, with emission cores at the centre of their Balmer absorption lines, 
probably indicating a close low-mass companion which is not detected in the continuum.
We classify these 3 sources as DAe white dwarfs (Fig.\,8 of Silvestri et al., 2006), consisting of 
a hot white dwarf with a very late M-dwarf companion. The \IPHAS photometry in Fig.\ \ref{fig:IPHASCCD} already shows 
that these white dwarfs have a companion and infrared colours can be used to determine the nature of the low-mass 
secondaries (Verbeek et al., in prep.). 
An other option is that these 3 systems are Dwarf Novae (DN) (Aungwerojwit et al., 2005, Morales-Rueda \& Marsh, 2002),
or they might be `pre-CV' candidates (Tappert et al., 2009 and Szkody et al., 2007).
Although the hydrogen emission lines of these 3 systems is narrow, the helium emission lines
are only possible when there is accretion, so they could be Cataclysmic Variables.
Other objects in the UV-excess spectra are 2 Quasi Stellar Objects (QSOs), 
one with redshift $z$=2.16 at $(l,b)$=(125\degr.44, --4\degr.29)
and one with redshift $z$=1.48 at $(l,b)$=(117\degr.29, --4\degr.43),
classified using example spectra (Fig.\,7 of Brunzendorf et al., 2002).
For UVEXJ0110+5829 the bluest broad line in the spectrum is Ly$\alpha$
and the emission line at 6\,030 \AA\, is CIII.
In the spectrum of UVEXJ0008+5758 the bluest line is CIV,
the second line is CIII and the emission line at $\sim$7\,000 \AA\, is MgII.\\

There are two more UV-excess sources (UVEXJ2026+4050 and UVEXJ2049+3811) both with a Hectospec spectrum, 
showing a continuum with several emission lines at the position of the Balmer lines. 
These emission features are probably not real since they are also present in the offset sky spectrum.
Their photometry $(r-H\alpha)$$\sim$0.4 confirms the emission features in the spectra.
The narrow Balmer and HeI emission lines, indicating a low-density environment, are mostly (or completely) from the huge diffuse
emission in the field, likely a HII region. Without the emission lines they
might just be hot white dwarfs. Based on the available spectra the sources can 
not be classified, so they are labeled `unknown'.\\

For two other UV-excess sources (UVEXJ0110+6004 and UVEXJ2047+4155) there are 
Calar Alto 2.2m spectra available. These UV-excess sources are not included in this paper since 
they are the known Cataclysmic Variables `HT Cassiopeiae' (Rafanelli, 1979) and `V516 Cygni' (Spogli et al., 1998).\\

\section{Discussion and conclusions}
\label{sec:discussion}
The main conclusion is that of the sources in the UV-excess catalogue 95$\%$ are genuine UV-excess sources,
such as white dwarfs, white dwarf binaries, subdwarf stars type O and B and QSOs.
Five percent of the UV-excess candidates are classified as main-sequence (MS) stars or blue horizontal branch (BHB) stars with spectral types later than A0V.
If the sources classified as MS/BHB are main-sequence stars, the fitting of the Pickles library spectra (Pickles, 1998) shows that 4 sources 
are slightly reddened F0V stars with reddening $E(B-V)$$\leq$0.1 and 3 sources are B0V-B3V stars with reddening 0.4$\leq$$E(B-V)$$\leq$0.5.
Their spectra look like B-type main-sequence stars, but the Balmer absorption lines are stronger than the Balmer lines of the best fit template spectra.
Since gravity is slightly high they could be horizontal branch stars, although usually BHB stars have less helium absorption.
Two of the B-type MS/BHB stars slightly blue shifted so might be high velocity stars, UVEXJ1940+3220 has 
a velocity of $RV$=320km/s and UVEXJ1938+3054 has a velocity of $RV$=163km/s.
There is 1 G-type star with as best fit a Pickles G2V star with reddening $E(B-V)$$\sim$0.3.
G-type stars are expected to have colours redder than $(g-r)$$>$0.6 and $(U-g)$$>$0.4, but they can enter the UV-excess
region when they are metal weak, i.e. subdwarfs type with less light blocked at the blue/UV wavelengths (Eracleous et al., 2002).
Since the number of late type stars in the fields is large, photometric errors can cause a few outliers to be scattered into the
UV-excess selection region (Krzesinski et al., 2004).\\

Secondly, in the colour-colour and colour-magnitude diagrams of Figs.\,6 and 7 of V12 about 20$\%$ of the UV-excess 
sources overlaps with the location of the `subdwarfs' population at $(g-r)$$>$0.3 and $(U-g)$$>$0.2. 
These UV-excess sources that overlap the subdwarf area in the colour-colour and colour-magnitude diagrams 
are  2 QSOs, 7 Cataclysmic Variables, 1 DAe, 2 T Tauri stars, 1 Be star, 3 DA+dM stars,
3 He-sdO stars, 1 sdO star, 2 sdB stars, 1 B-type MS/BHB star, 3 F-type MS/BHB stars and 1 DAB white dwarf. 
For our selection of UV-excess candidates from the \UVEX data this means that
when the aim is to find white dwarfs, a colour cut can be applied to decrease the number of other objects.
But leaving these sources at the location of the `subdwarfs' out from the UV-excess catalogue would lead to a loss of most QSOs, 
Cataclysmic Variables, T Tauri stars, Be stars and DA+dM stars.\\

The location of the different populations in the $(U-g)$ vs. $(g-r)$ colour-colour diagram and the 
$g$ vs. $(U-g)$ and $g$ vs. $(g-r)$ colour-magnitude diagrams is shown in Fig.\ \ref{fig:classCCD} and\ \ref{fig:classCMD}. 
Their positions match with the positions of the populations in the colour-colour diagrams of other surveys (e.g. Fig.\,1 of Krzesinski et al., 2004, 
Fig.\,1 of Harris et al., 2003, Fig.\,3 of Stobie et al., 1997, Fig.\,3 of Yanny et al., 2009 and the Figs. of Kilkenny et al., 1997).
The locations of the classified sources in the colour-colour and colour-magnitude diagrams agree with the locations of the sources 
with a Simbad match in the colour-colour and colour-magnitude diagrams in Fig.\,9 of V12.
There is a clear relation between the different kind of sources and the way they are selected in V12. 
The way the sources were selected from the colour-colour and colour-magnitude diagrams is captured in the `selection label' 
(column 20 of the UV-excess catalogue, Appendix A of V12), and is summarized for our classified sources in Table.\ \ref{tab:selection}.
Only 2 DA white dwarfs were selected less than 0.4 magnitude from the blue edge in the $g$ vs. $(g-r)$ colour-magnitude diagram, the other 60 DA
white dwarfs were selected more than 0.4 magnitude from the blue edge in the $g$ vs. $(g-r)$ colour-magnitude diagram. All DB and DC white dwarfs
were selected both were selected more than 0.4 magnitude from the blue edge in both colour-magnitude diagrams 
and in the $(U-g)$ vs. $(g-r)$ colour-colour diagram,
while most DBA white dwarfs were selected less than 0.4 magnitude from the blue edge in the $g$ vs. $(g-r)$ colour-magnitude diagram 
and in the $(U-g)$ vs. $(g-r)$ colour-colour diagram.
The 2 QSOs and the majority of the H$\alpha$ emission line objects were selected in the $g$ vs. $(U-g)$ colour-magnitude diagram 
but not in the $g$ vs. $(g-r)$ colour-magnitude diagram. 
We could improve the selection method of V12 using these spectroscopic results by taking only sources in the UV-excess 
catalogue with favourable selection labels into account. This will increase the number of genuine UV-excess objects to 97$\%$ by 
e.g. leaving out selection labels `514' and `518' since the largest fraction MS/BHB stars have these selection labels, but this will also lead
to a loss of some peculiar objects such as DAe stars, Cataclysmic Variables and Be stars.
The \UVEX and \IPHAS photometry can also be combined with other (infrared and ultraviolet) surveys in order to improve the 
selection of different populations (Verbeek et al., in prep.).\\

\begin{table*}
\caption[]{The selection in V12 of the classified UV-excess spectra. \label{tab:selection} }
{\small
\begin{tabular}{ | l | l | l | }
    \hline
Label    &  Selected from &  Objects \\ \hline
514	     &  $g$ vs. $(U-g)$  						 	&  1DAe, 1Be, 1MS/BHB, 1He-sdO, 1sdB+F  \\
515          &  $g$ vs. $(U-g)$ \& $(U-g)$ vs. $(g-r)$			 		&  2CV, 2QSO, 1TT, 1DAB+dM, 1He-sdO, 1BHB/MS  \\
518	     &  $g$ vs. $(U-g)$ \& $<$0.4$g$ vs. $(g-r)$			 	&  1CV, 1DAe, 2MS/BHB, 1noisy  \\
519	     &  $g$ vs. $(U-g)$ \& $<$0.4$g$ vs. $(g-r)$ \& $(U-g)$ vs. $(g-r)$	 	&  5CV, 4DBA, 2DA, 1TT, 2DA+dM, 4sdB, 2He-sdO, 1BHB/MS, 1G, 4noisy   \\

1028	     &  $g$ vs. $(g-r)$  						 	&  28DA, 1DZA, 1unknown, 1DA+dM, 1MS/BHB, 1sdB, 1noisy   \\
1029	     &  $g$ vs. $(g-r)$ \& $(U-g)$ vs. $(g-r)$			 		&  3DA, 1DA+dM   \\
1031         &  $g$ vs. $(g-r)$ \& $<$0.4$g$ vs. $(U-g)$ \& $(U-g)$ vs. $(g-r)$	 	&  1DA, 1MIII   \\
1542         &  $g$ vs. $(g-r)$ \& $g$ vs. $(U-g)$					&  1unknown   \\
1543	     &  $g$ vs. $(g-r)$ \& $g$ vs. $(U-g)$ \& $(U-g)$ vs. $(g-r)$  		&  28DA, 4DB, 1DBA, 4DC, 1DAe, 1DZ, 1sdO, 1BHB/MS, 6noisy   \\
    \hline
\end{tabular} \\ 
}
\end{table*}

About 64$\%$ of the UV-excess candidates turn out to be white dwarfs.
The fitting of the white dwarf models to the UV-excess hydrogen atmosphere white dwarf spectra shows a
distribution of 9\,000K$<$$T_{\rm eff}$$<$65\,000K and an average surface gravity of $log\,g$$\sim$8. 
These results are in agreement with the results of other studies. (Liebert et al., 2005, Bergeron et al., 1992,
Napiwotzki et al., 1999, Finley et al., 1997, Gianninas et al., 2011 and Kepler et al., 2007).
The accuracy of the continuum fitting method applied in Sect.\ \ref{sec:whitedwarf} depends 
on the SNR and the flux calibration of the spectra.
For white dwarfs the accuracy of the temperature fit will approach the surface temperature to typically $\sim$1\,000$K$ for white dwarfs 
with T$<$20\,000 and $\sim$2\,000$K$ for the hotter white dwarfs with T$\geq$20\,000, for spectra with signal to noise SNR$>$20.\\

When we extrapolate the result that 64$\%$ of the UV-excess catalogue sources are white dwarfs, 
the complete \UVEX survey will bring up a sample of $\sim 1.2 \times 10^{4}$ new white dwarfs 
($\sim$7 per square degree). If we only look at UV-excess white dwarf sample brighter than $g$$<$20, 
\UVEX will bring up a sample of $\sim$4000 new white dwarfs with $g$$<$20 in the full survey area.
The UV-excess sample might not be complete for the coolest white dwarfs below $T$$<$10\,000K since they have too red colours. There is also the additional 
problem of dust extinction (Sale et al., 2009), which has only a small effect on the local white dwarf sample while it merely screens out more distant objects.
As shown in Sect.\ \ref{sec:whitedwarf} reddening is typically $E(B-V)$$\leq$0.1 magnitudes 
for most of the white dwarfs in the UV-excess catalogue. A space density of white dwarfs (Holberg et al., 2008) 
in the Galactic Plane and a comparison with population synthesis predictions will be further discussed in Verbeek et al., (in prep.).\\ 
 
\begin{figure}
\centerline{\epsfig{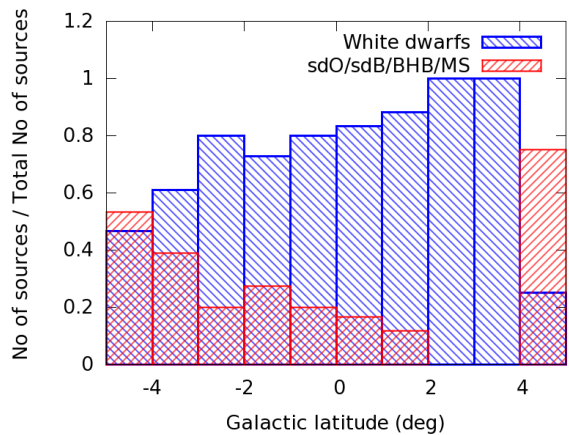}}
\caption{Galactic latitude distribution of the sources classified as white dwarfs (blue) 
and the sources classified as sdO/sdB stars, MS/BHB stars and ``noisy'' (red). The number of sources per bin is 
normalized by the total number of obtained spectra in the latitude bin. 
\label{fig:latitude}}
\end{figure}
 
The Galactic latitude distribution of the sources classified as white dwarfs and 
as sdO/sdB stars, main-sequence stars and blue horizontal branch stars is shown in Fig.\ \ref{fig:latitude}.
The sources labeled as ``noisy'' in Sect.\ \ref{sec:mainseqsubdwarf} are add to the sdO/sdB/BHB/MS sample since they
probably are sdO/sdB or MS/BHB stars.
The white dwarfs are mainly detected at Galactic latitudes smaller than $|b|$$<$4, while 
the distribution of sdO/sdB stars and MS/BHB stars peaks at Galactic latitudes larger than $|b|$$>$4.
This result can be explained by the absolute magnitude distribution of the different populations in combination with the effect of extinction, 
as can be seen in Fig.\,1 of Groot et al. (2009).\\

\begin{figure}
\centerline{\epsfig{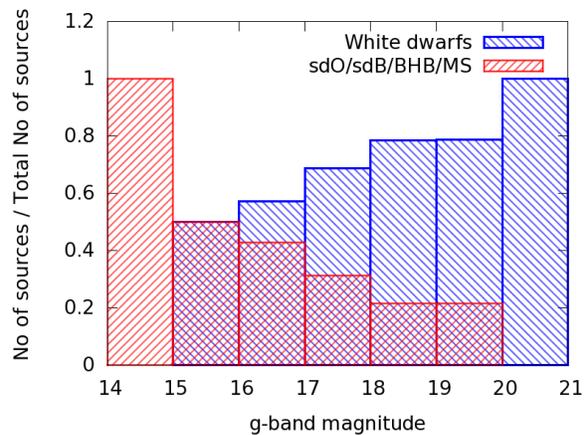}}
\caption{Magnitude distribution of the sources classified as white dwarfs (blue) 
and the sources classified as sdO/sdB stars, MS/BHB stars and ``noisy'' (red). The number of sources per bin is 
normalized by the total number of obtained spectra in the magnitude bin.
\label{fig:magnitude}}
\end{figure}

The magnitude distribution of the spectra classified as white dwarfs and as sdO/sdB stars, MS/BHB stars and ``noisy'' is shown in Fig.\ \ref{fig:magnitude}.
The fraction of white dwarfs clearly increases for fainter $g$-band magnitudes.
The total number of white dwarfs increases strongly for fainter magnitudes since also the number of selected sources increases 
(see e.g. Fig.\,7 of V12 and Fig.\,1 of Bergeron et al., 1992).
The fraction of MS/BHB and subdwarf sources is larger for the brighter $g$-band magnitudes, even with the sources classified as ``noisy'' 
included in the sdO/sdB/BHB/MS sample. 
A larger fraction of white dwarfs is found at magnitudes fainter than $g$$>$17.\\

\begin{figure*}
\centerline{\epsfig{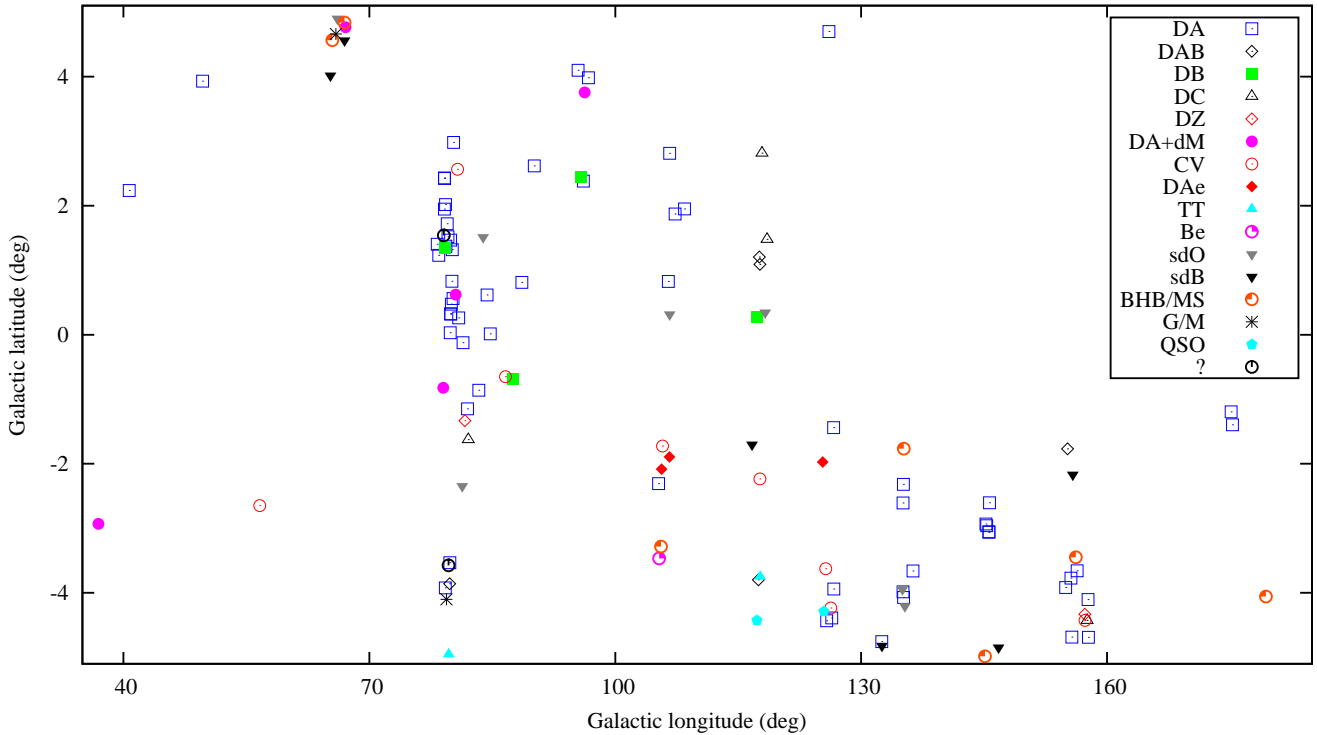}}
\caption{Galactic latitude vs. Galactic longitude diagram with all obtained UV-excess spectra.
The classified sources are indicated with the symbols of Figs.\,1 to 3.
\label{fig:LatitudeVSLongitude}}
\end{figure*}

If we assume that UV-excess candidates classified as main-sequence stars and blue horizontal branch stars are all MS stars,
we can estimate the distance $d$ using $d$=$10^{0.2\times(m-M-A_{V})}\,10pc$,
where we use the observed $g$-band magnitude as apparent magnitude ($m$), $M$ is the absolute magnitude and
$A(V)$ is the total extinction for the $V$-band filter. 
Since $A(V)$ = $R_{V}$$\times$$E(B-V)$, where we use $R_{V}$=3.1 for the indicator of dust grain size distribution and 
the results of the fitting in Sect.\ \ref{sec:mainseqsubdwarf} for the reddening $E(B-V)$=$A(B)-A(V)$, 
we can estimate a distance range per source. 
In our UV-excess sample the F-type MS/BHB stars have a $g$-band magnitude of 17.3$<$$g$$<$19.6 and reddening $E(B-V)$=0.1,
the G2V star has a $g$-band magnitude of $g$=19.5 and $E(B-V)$=0.3, 
the B-type BHB/MS stars have a typical $g$-band magnitude between 14.5$<$$g$$<$18.6 and $E(B-V)$=0.4,
Taking into account the effect of reddening the distance estimations would be 
$\sim$5kpc for the G2V star with $g$=19.5 and $\sim$8kpc for the F0V star with $g$=17.3 which is within the Milky Way.
For the fainter F-type and B-type stars the distances would be 
$\sim$20kpc for the F0V star with $g$=19.6 and
$\sim$35kpc for the B0V star with $g$=17.0
if they would be main-sequence stars.
These distances would be outside the Milky Way. Since their colours are only slightly reddened they must be intrinsically fainter objects.
So, we conclude that these objects must be blue horizontal branch stars or subdwarf type stars.\\

An interesting side benefit is the detection of the 2 broadline QSOs in the UV-excess spectra, with $z$$\sim$2.16 and $z$$\sim$1.48 and at $|b|$=4.
Only about $\sim$10 QSOs are found at low Galactic latitude regions (Im et al., 2007, Lee et al., 2008 and Becker et al., 1990).
The Schlegel map (Schlegel et al., 1998) gives a reddening of $E(B-V)$$\sim$0.5 for both QSOs.
Due to the internal reddening of the QSOs we can not directly estimate the amount of reddening caused by our 
Milky Way from their spectra (Knigge et al., 2008).\\

Of the UV-excess spectra 122 have a match in \IPHAS. 
These matches are shown in the colour-colour and 
colour-magnitude diagrams of Fig.\ \ref{fig:IPHASCCD}. Nine of the classified UV-excess sources are in the Deacon IPHAS-POSSI PM 
catalogue (Deacon et al., 2009): 7 DA white dwarfs, 1 DC white dwarf and 1 DA+dM binary system.
Except for the DA+dM at $(r-i)$=1.3 all sources overlap with the location of the white dwarf population at $(r-i)$$\sim$0 in Fig.\,11 of V12.
Eight of the classified UV-excess sources are in the Witham H$\alpha$ emission line catalogue (Witham et al., 2008): 
4 Cataclysmic Variables, 1 Be star, 1 Classical T Tauri star and 2 sdO candidates. 
There are some sources with clear H$\alpha$ emission lines in their spectra that are not 
in the Witham H$\alpha$ emission line catalogue. Four sources classified as Cataclysmic Variables clearly show H$\alpha$
emission in the \IPHAS colour-colour diagram of Fig.\ \ref{fig:IPHASCCD}. Two of these Cataclysmic Variables are not 
in the Witham catalogue because they have $r$-band magnitudes $r$$>$19.5. 
The H$\alpha$ emission of some other Cataclysmic Variables with $EW$$<$20\AA\, is probably not 
strong enough to be in the Witham catalogue, or they can also have variable emission.\\

\subsection{Comparison with spectroscopic surveys}
\label{sec:crossmatch}

\begin{itemize}

\item We can compare our results with the spectroscopic observations of Eracleous et al. (2002) of
27 UV-bright stars, with $(U-B)$$<$--0.2 and magnitude 13$<$$B$$<$16, selected from the 
Sandage Two-colour Galactic Plane survey, in the Lanning catalogue (Lanning, 1973).
This sample contains 2 DA white dwarfs, 1 DB white dwarfs, 1 DA+dM, 16 O/B stars (60$\%$), 1 F/G star, 1 M star, 1 sdO, 2 subdwarfs, 
1 composite object and 1 emission line star. When we compare this sample with the sources in our UV-excess sample the distribution of spectral types 
is similar. The fraction of white dwarfs and O/B stars is very different for both surveys, 
which might be due to the magnitude limit 13$<$$B$$<$16 of the Sandage survey and the criteria used for the classification.\\

\item 
A second sample of 46 UV-bright sources from the Sandage Two-color Survey obtained by L\'epine et al. (2011)
contains 29 DA white dwarfs (63$\%$), 5 DB white dwarfs (11$\%$), 3 DC white dwarfs, 1 DZ white dwarf, 1 DA+dM, 1 sdB, 2 sdO and 
4 F-type stars. Here the F-type stars are at Galactic latitudes larger than $|b|>$5.
When we compare this sample with the sources in the UV-excess sample the distribution of spectral types and their variety is very similar,  
e.g. fraction of white dwarfs. The number of H$\alpha$ emission line objects in the Sandage survey is very different from our UV-excess sample.\\

\item The Kitt Peak-Downes (KPD) survey (Downes et al., 1986) sample of 158 UV-excess objects at
Galactic latitude $|b|$$<$12\degr, brighter than $B$$<$15.3 and $(U-B)$$<$--0.5 contains
21 DA white dwarfs, 13 white dwarfs of other types, 20 sdO, 40 sdB, 5 Planetary Nebulae, 41 Be stars, 9 Cataclysmic Variables and 9 other peculiar sources.
Remarkable is the small fraction of white dwarfs (only $22\%$) and the high number of Be stars and Planetary Nebulae 
in the KPD survey compared to the \UVEX survey. This might be partly due to the Galactic latitude difference of the two surveys 
and the fact that the detection of Planetary Nebulae is strongly affected by interstellar obscuration 
(Fig.\,6 of Miszalski et al., 2008, Fig.\,7 of Parker et al., 2006 and Moe and De Marco, 2006) at Galactic latitudes smaller than $|b|$$<$5.
Normally narrow-band and red/IR surveys would be required to select new Planetary Nebulae. 
The low number of main-sequence sources in the KPD survey can be explained by the demand $(U-B)$$<$--0.5 and their classification of all blue continuum 
spectra with strong Balmer lines as sdB candidates. This also directly explains why the number of sdO and sdB stars in KPD is 
reversed compared to \UVEX. Despite the different magnitude depths and colour cuts the fraction of e.g. QSOs, 
Cataclysmic Variables and DC white dwarfs is the same for both surveys.
The distribution of different spectral types over Galactic latitude and Galactic longitude varies strongly as can be seen in Fig.\ \ref{fig:LatitudeVSLongitude}.
When we compare only the KPD sources at Galactic latitude smaller than $|b|<$5 and do not take the H$\alpha$ emitters into account, the
result is 16 DA, 2 DB, 1 DC, 13 sdB and 4 sdO stars. This result is similar to our classified UV-excess spectra.\\

\end{itemize}

\section*{Acknowledgement}
This paper makes use of data collected at the Isaac Newton Telescope, operated on the island of La Palma by the Isaac Newton Group in the
Spanish Observatorio del Roque de los Muchachos of the Inst\'{\i}tuto de Astrof\'{\i}sica de Canarias.
The observations were processed by the Cambridge Astronomy Survey Unit
(CASU) at the Institute of Astronomy, University of Cambridge.
Hectospec observations shown in this paper were obtained at the MMT Observatory, a joint facility of the University of Arizona and the Smithsonian Institution.
The \IPHAS FAST spectra shown in this work were obtained at Fred Lawrence Whipple Observatory (FLWO) on Mount Hopkins, Arizona.
We gratefully acknowledge the \IPHAS consortium for making available the Hectospec and FAST observations.
This research has made use of the Simbad database and the VizieR catalogue access 
tool, operated at CDS, Strasbourg, France.
KV is supported by a NWO-EW grant 614.000.601 to PJG. 
The authors would like to thank Detlev Koester and Pierre Bergeron for
making available their white dwarf model spectra.\\

\appendix

\section{List of obtained UV-excess spectra}
\label{app:appendix}
All UV-excess spectra, their features and classification are summarized in Table\ \ref{tab:spectra}, sorted by right ascension.
The columns contain the spectrum number and name composed of the \UVEX right ascension and declination, the Galactic longitude and latitude,  
the \UVEX field in which the object was selected, the \UVEX selection label (column 20 of the UV-excess catalogue, described in V12), 
the \UVEX photometry, if available the \IPHAS photometry, 
the observing run and the classification. 
The classification is summarized in the two last columns. Column 14 shows the type of object and the ``by eye'' most appropriate spectral 
type of the spectrum. Sources which have no good fitting result are labeled as ``noisy'' in column 14. 
Column 15 shows the result of the spectra fitting: for white dwarfs the effective temperature $T_{\rm eff}$ in kK
and if available the surface gravity $log\,g$. For He-sdO, sdO, sdB and BHB/MS B-type sources column 15 shows
$T_{\rm eff}$ in kK, $log\,g$ and $log\,(n(He)/n(H))$ and for 
F-type MS/BHB sources without He lines $T_{\rm eff}$ in kK and $log\,g$. 
For sources classified as MS/BHB and for most sources labeled as ``noisy'' column 15 shows the most appropriate main-sequence spectral 
type and reddening $E(B-V)$. For the H$\alpha$ emission line stars column 15 shows the $FWHM$ and $EW$ of the $H\alpha$ line
given in units of \AA.\\

\begin{figure*}
\centerline{\epsfig{file=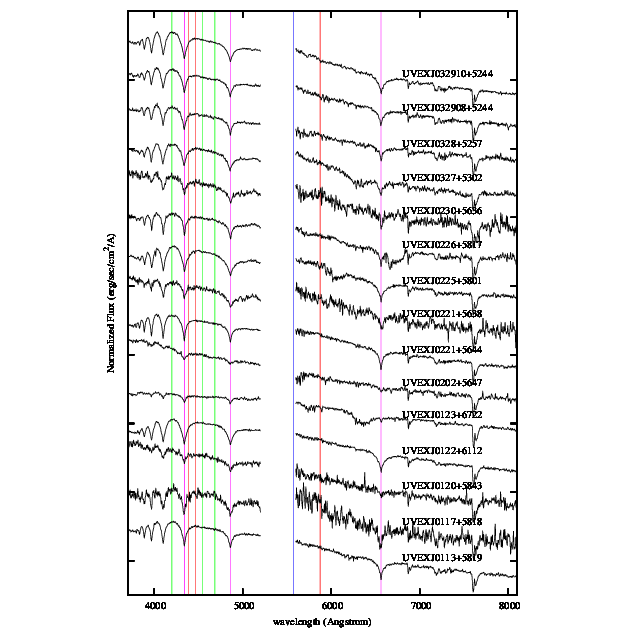,width=22cm,angle=0,clip=}}
\caption{DA white dwarfs. The pink, red and green lines indicate the position of hydrogen, HeI and HeII respectively. The blue line is a skyline in the
Hectospec spectra.
\label{fig:spectra1}}
\end{figure*}

All obtained UV-excess spectra are shown in Figs.\ \ref{fig:spectra1} to\ \ref{fig:spectra11} per population sorted by right ascension. 
These spectra and Table\ \ref{tab:spectra} can also be obtained from the \UVEX website. Note that the Hectospec spectra might show emission features at 
the wavelengths of the Balmer lines due to bad sky subtraction in fields with diffuse emission (Fabricant et al., 2005). 
The Hectospec spectra were corrected for incomplete sky subtraction (Vink et al., 2008), and they are not flux calibrated. 
Some of the WHT/ISIS spectra obtained in December 2010 have a small hump around 5\,100\AA. which is not a real feature. 
Also note the dichroic gap from $\lambda$=5\,200--5\,600\AA,\ of the WHT/ISIS spectra due to the blue and red arm of ISIS. 
All UV-excess spectra were smoothed by a boxcar smoothing algorithm which takes for each pixel also the flux of four neighbouring pixels into account with
weights 1:2:4:2:1.\\

\subsection{Notes to individual objects}
\label{sec:notes}

\begin{itemize}

\item UVEXJ001102.23+584232.3: Classified as T Tauri candidate with underlying M4V atmosphere. CaII emission and strong hydrogen lines where $H\alpha$ has  
 $FWHM$=6\AA\, and $EW$=--80\AA \\

 \item UVEXJ011053.07+604830.9: Classified as DAe white dwarf, showing narrow H$\alpha$, H$\beta$ and H$\gamma$ emission lines in broad 
 absorption lines with additionally HeI emission, so could also be a Dwarf Nova. No sign of a companion.\\

 \item UVEXJ012359.82+672223.1: DA white dwarf. The bump at $\lambda$=6\,300\AA\, is due to the data calibration.\\

 \item UVEXJ041926.84+440058.4: DC white dwarf, showing a very blue continuum spectrum and possibly some weak HeI absorption.\\

 \item UVEXJ041045.70+461137.1: Only a blue WHT/ISIS spectrum was obtained for this source, sufficient to clearly classify the source as a DA white dwarf.\\

 \item UVEXJ041840.30+441714.1: DZ white dwarf showing a continuum with clear CaII H and K absorption in combination with weak hydrogen and HeI absorption lines, 
 similar to the DZ spectra of Sion et al. (1990).\\

 \item UVEXJ202457.34+410804.1: There is a gap in the red spectrum at $\lambda$=7\,500\AA. The absorption features at $\lambda$=6\,200\AA\, are not real.\\

 \item UVEXJ202630.19+405024.1: Classified as `unknown'. The Balmer and HeI emission lines in the spectrum of this source are also present in the sky 
 offset spectrum due to diffuse emission in the field.\\

 \item UVEXJ202712.06+424720.1: Novalike CV with broad, double peaked Balmer and Helium lines.\\

 \item UVEXJ202744.63+405044.5: DB white dwarf. The H$\alpha$ and H$\beta$ emission lines are not real.\\ 

 \item UVEXJ202800.47+405620.0: H$\alpha$ nebula structure around object in \IPHAS finders.\\

 \item UVEXJ203656.54+392934.9: Classified as `noisy', has a white dwarf type spectrum showing several odd lines.
 There is a nearby red star on the \IPHAS images.\\

 \item UVEXJ204649.51+410906.2: DZ white dwarf with clear CaII H\&K and no other lines. The H$\alpha$ line is not real.\\

 \item UVEXJ204710.61+413133.5: Unclear features in the red part of the spectrum, which may be due to a red companion. 
 Remarkable: this source is the bluest DA in Fig.\,1 while it has $(HeI-r)$=0.23.\\

 \item UVEXJ204751.27+442920.1: Odd shaped Balmer lines and unclear features in the red part of the spectrum, which may be due to a red companion.\\
 
 \item UVEXJ204923.48+381139.0: Classified as `unknown'. The emission lines are also present in the sky offset spectrum. 
 These lines are mostly (or completely) due to huge diffuse emission in the field. 
 The sources has a match in the IPHAS-POSS PM catalogue (Deacon et al., 2009). 
 This indicates that the sources must be an evolved stellar objects within the Milky Way.\\

 \item UVEXJ204945.83+382057.2: DA white dwarf. Red part of the spectrum too noisy (faint source $g$=19.7).\\

 \item UVEXJ205039.07+373958.4: M-giant with type M5III showing clear TiO bands. No sign of a companion is found in the blue part of the spectrum.
 Without a white dwarf companion normally these type of objects are found at $(g-r)$$\sim$1.5.\\

 \item UVEXJ205449.65+371953.2: T Tauri candidate with underlying M5V atmosphere. Weak hydrogen lines where $H\alpha$ has $FWHM$=4\AA\, and $EW$=--9\AA\,, 
 due to low mass accretion (Fig.1 of Barentsen et al., 2012).\\

 \item UVEXJ224112.21+564419.1: CV with broad Balmer and Helium lines, including He II 4686: could be magnetic.\\

 \item UVEXJ224145.94+562230.0: Classified as DAe, could be a Dwarf Nova, showing narrow H$\alpha$ emission and broad Balmer absorption lines.\\

\end{itemize}

\oddsidemargin=-2cm
\evensidemargin=-2cm

\renewcommand*\thetable{\Alph{section}A\arabic{table}}

\begin{table*}
\caption[]{Classifications of the UV-excess spectra. \label{tab:spectra} }
\centering
\begin{sideways}
\tiny
\noindent
\begin{tabular}{p{3mm}@{\ \ \ }p{30mm}@{\ \ \ }p{10mm}@{\ \ \ }p{10mm}@{\ \ \ }p{6mm}@{\ \ \ }p{8mm}@{\ \ \ }p{7mm}@{\ \ \ }
p{7mm}@{\ \ \ }p{7mm}@{\ \ \ }p{7mm}@{\ \ \ }p{8mm}@{\ \ \ }p{10mm}@{\ \ \ }p{17mm}@{\ \ \ }p{16mm}@{\ \ \ }p{16mm}}
\hline 
No & Name & l & b & Field & Selection & $g$ & ($U-g$) & ($g-r$) & ($HeI$) & ($r-i$) & ($r-H\alpha$) & Run & Classification & Fit \\ \hline
1   & UVEXJ000016.27+603246.3  &  116.67893 &  -1.70286   & 9 & 519 & 16.761 & -0.687 & 0.212 & 16.534         & -0.433  & 0.371    &  WHT2010dec12	   &	 sdB		& 34.4/6.0/--2.8 \\
2   & UVEXJ000218.56+623649.3  &  117.31640 &  +0.27881   & 6 & 1543 & 18.073 & -0.852 & 0.017 & 18.282        & 0.001	 & 0.281    &  WHT2010dec12	   &	 DB		& 16.0 \\
3   & UVEXJ000310.20+633430.2  &  117.59252 &  +1.20480   & 8 & 519 & 17.387 & -0.796 & 0.07 & 17.492	       & -0.037  & 0.531    &  WHT2010dec12	   &	 DBA		& 14.0 \\
4   & UVEXJ000313.11+651248.1  &  117.90273 &  +2.81352   & 22 & 1543 & 17.619 & -0.584 & 0.149 & 17.516       & 0.263	 & 0.177    &  WHT2008oct03	  &	DC		&  \\
5   & UVEXJ000355.86+632833.2  &  117.65757 &  +1.09168   & 19 & 519 & 16.380 & -0.583 & 0.149 & 16.232        & 0.077	 & 0.062    &  Fast2009feb03	  &	DAB		&  \\
6   & UVEXJ000843.60+601154.7  &  117.64423 &  -2.23399   & 40 & 515 & 19.837 & -0.485 & 1.179 & 18.649        & 0.879	 & 0.829    &  WHT2010dec13	   &	 CV		& 18/--50 \\
7   & UVEXJ000848.64+575832.7  &  117.28742 &  -4.42807   & 48 & 515 & 19.494 & -0.092 & 0.869 & 18.743        & 0.734	 & 0.360    &  WHT2010dec12	   &	 QSO		& (z=1.48) \\
8   & UVEXJ000919.43+583729.7  &  117.46085 &  -3.79876   & 52 & 519 & 18.162 & -0.217 & 0.729 & 17.597        & 0.563	 & 0.238    &  WHT2010dec12	   &	 DAB		&  \\
9   & UVEXJ000945.33+631032.4  &  118.24989 &  +0.68346   & 51 & 1543 & 17.983 & -0.962 & -0.041 & 17.967      & 0.074	 & 0.185    &  WHT2010dec14	   &	 noisy         & O5V,0.3 \\
10  & UVEXJ001032.27+625050.0  &  118.28513 &  +0.34507   & 62 &  514 & 16.427 &  0.108 &  0.555 & 15.901      & 0.429	 & 0.233    &  Fast2011nov30	   &	 He-sdO    &   \\
11  & UVEXJ001101.26+640013.7  &  118.52021 &  +1.47880   & 55 & 1543 & 19.564 & -0.476 & 0.306 & 19.360       & 0.349	 & 0.189    &  WHT2009sep21	   &	 DC		&  \\
12  & UVEXJ001102.23+584232.3  &  117.69466 &  -3.75135   & 52 & 515 & 19.710 & -0.151 & 1.548 & 18.743        & 1.932	 & 1.381    &  WHT2010dec10	   &	 TTauri(M4V)	& 6/--80 \\
13  & UVEXJ011037.91+582928.1  &  125.44465 &  -4.28694   & 404 & 515 & 19.518 & -0.158 & 0.697 & 19.003       & 0.576	 & 0.403    &  WHT2009sep21	   &	 QSO		& (z=2.16) \\
14  & UVEXJ011053.07+604830.9  &  125.30232 &  -1.97363   & 410 & 1543 & 17.026 & -0.895 & -0.006 & 17.162     & 0.128	 & 0.168    &  WHT2010dec13	   &	 DAe	 &  \\
15  & UVEXJ011102.67+594020.8  &  125.40836 &  -3.10500   & 389 & 1543 & 18.839 & -0.561 & 0.143 & 18.672      & -0.104  & 0.218    &  WHT2009sep22	   &	 noisy       &  \\
16  & UVEXJ011245.47+590757.3  &  125.66906 &  -3.62581   & 405 & 519 & 18.306 & -0.272 & 0.598 & 17.681       & 0.298	 & 0.668    &  WHT2009sep23	   &	 CV		& 25/--150 \\
17  & UVEXJ011311.87+581902.3  &  125.79454 &  -4.43345   & 402 & 1028 & 18.130 & 0.052 & 0.153 & 18.052       & 0.095	 & -0.125   &  WHT2009sep21	   &	 DA		& 10.0 \\
18  & UVEXJ011712.36+582804.4  &  126.30500 &  -4.23500   & 436 & 515 & 19.143 & -0.798 & 0.906 & 0.000        & 0.857	 & 0.892    &  WHT2010dec12	   &	 CV		& 14/--60 \\
19  & UVEXJ011754.90+581815.4  &  126.41435 &  -4.38833   & 436 & 1028 & 17.736 & -0.31 & -0.031 & 0.000       & 0.005	 & -0.231   &  WHT2010dec14	   &	 DA		& 11.9/8.3 \\
20  & UVEXJ012015.68+584318.2  &  126.67481 &  -3.94040   & 437 & 1028 & 18.838 & -0.761 & 0.039 & 0.000       & 0.119	 & 0.094    &  WHT2010dec13	   &	 DA		& 36.9/8.1 \\
21  & UVEXJ012219.81+611229.9  &  126.64407 &  -1.44059   & 463 & 1543 & 17.509 & -0.318 & -0.017 & 0.000      & -0.015  & -0.136   &  WHT2009sep25	   &	 DA		& 12.8/8.1 \\
22  & UVEXJ012359.82+672223.1  &  126.06509 &  +4.69962   & 457 & 1543 & 18.659 & -0.598 & 0.175 & 0.000       & 0.050	 & 0.108    &  WHT2009sep25	   &	 DA		& 57.0 \\
23  & UVEXJ020201.82+564744.8  &  132.52194 &  -4.75581   & 679 & 1543 & 19.230 & -0.948 & -0.049 & 19.372     & no data & no data  &  WHT2009sep23	   &	 DA		& 65.0 \\
24  & UVEXJ020201.85+564342.3  &  132.54031 &  -4.82064   & 679 & 519 & 15.238 & -0.667 & -0.019 & 15.319      & no data & no data  &  WHT2009sep25	   &	 sdB	    & 27.5/5.5/--2.8 \\
25  & UVEXJ022113.52+564810.7  &  135.02814 &  -3.94521   & 784 & 519 & 17.267 & -0.833 & 0.375 & 16.989       & -0.026  & 0.162    &  WHT2009sep24	   &	 He-sdO 	& 44.0/5.5/1.4 \\
26  & UVEXJ022135.47+564436.6  &  135.09570 &  -3.98401   & 784 & 519 & 18.582 & -0.318 & 0.383 & 18.200       & -0.080  & -0.084   &  WHT2009sep22	   &	 DA		& 10.1/8.0 \\
27  & UVEXJ022151.40+563815.7  &  135.16636 &  -4.07093   & 784 & 1543 & 19.606 & -0.892 & 0.199 & 19.379      & -0.327  & 0.199    &  WHT2010dec13	   &	 DA		& 28.6/8.2 \\
28  & UVEXJ022241.76+562702.2  &  135.33986 &  -4.20672   & 800 & 518 & 17.895 & -0.314 & 0.291 & 17.729       & 0.156	 & 0.115    &  WHT2009sep22	   &	 noisy        & B3V,0.4 \\
29  & UVEXJ022510.84+580156.6  &  135.10038 &  -2.60741   & 814 & 1543 & 17.920 & -0.118 & -0.49 & 0.000       & -0.030  & -0.223   &  WHT2010dec12	   &	 DA		& 13.1/8.0 \\
30  & UVEXJ022615.13+581710.2  &  135.14207 &  -2.31994   & 814 & 1028 & 18.970 & -0.251 & -0.167 & 0.000      & 0.108	 & -0.042   &  WHT2010dec12	   &	 DA		& 9.8/8.5 \\
31  & UVEXJ022815.18+584640.8  &  135.20774 &  -1.76697   & 820 & 514 & 18.200 & 0.106 & 0.959 & 0.000         & 0.525	 & 0.213    &  WHT2010dec12	   &	 MS/BHB        & 14.2/3.8/F0V,0.1 \\
32  & UVEXJ023044.92+563622.6  &  136.32179 &  -3.66150   & 838 & 1543 & 18.789 & -0.772 & -0.118 & 0.000      & -0.001  & 0.062    &  WHT2010dec13	   &	 DA		& 30.0/8.2 \\
33  & UVEXJ031943.45+512309.0  &  145.11863 &  -4.98344   & 1153 & 518 & 17.368 & 0.151 & 0.478 & 16.904       & 0.306	 & 0.169    &  WHT2010dec14	   &	 MS/BHB        & 15.8/4.3/F0V,0.0 \\
34  & UVEXJ032737.64+530231.1  &  145.21912 &  -2.93464   & 1206 & 1028 & 17.555 & -0.415 & -0.006 & 17.521    & -0.010  & -0.078   &  WHT2010dec14	   &	 DA		& 18.5/8.1 \\
35  & UVEXJ032807.05+525737.2  &  145.32615 &  -2.96070   & 1230 & 1543 & 18.372 & -0.558 & 0.06 & 18.432      & -0.147  & 0.044    &  WHT2009sep21	   &	 DA		& 23.9/7.8 \\
36  & UVEXJ032855.25+503529.8  &  146.76969 &  -4.84547   & 1224 & 519 & 14.202 & -0.558 & 0.127 & 14.085      & -0.015  & 0.109    &  WHT2009sep25	   &	 sdB		& 28.5/5.5/--2.5 \\
37  & UVEXJ032908.01+524400.6  &  145.58117 &  -3.06136   & 1222 & 1543 & 18.484 & -0.503 & 0.072 & 18.370     & 0.055	 & -0.067   &  WHT2009sep21	   &	 DA		& 22.3/7.8 \\
38  & UVEXJ032910.60+524426.3  &  145.58251 &  -3.05178   & 1222 & 1543 & 17.068 & -0.417 & -0.034 & 17.027    & -0.067  & -0.123   &  WHT2009sep21	   &	 DA		& 12.8/8.1 \\
39  & UVEXJ033118.06+530351.3  &  145.66179 &  -2.60337   & 1230 & 1543 & 16.975 & -0.505 & -0.055 & 16.919    & -0.198  & -0.196   &  WHT2009sep22	   &	 DA		& 17.1/8.1 \\
40  & UVEXJ041045.70+461137.1  &  154.95071 &  -3.91735   & 1567 & 519 & 17.596 & -0.307 & 0.098 & 17.447      & -0.038  & -0.135   &  WHT2010dec14	   &	 DA		& 14.5/7.6 \\
41  & UVEXJ041053.99+450706.5  &  155.70394 &  -4.68702   & 1553 & 1543 & 19.860 & -0.199 & 0.162 & 0.000      & 0.319	 & -0.375   &  WHT2009sep22	   &	 DA		& 16.1/7.8 \\
42  & UVEXJ041359.37+455151.2  &  155.58583 &  -3.77300   & 1585 & 1028 & 17.494 & -0.221 & 0.078 & 0.000      & -0.135  & -0.130   &  WHT2009sep21	   &	 DA		& 12.3/8.1 \\
43  & UVEXJ041733.05+452524.4  &  156.34147 &  -3.65845   & 1629 & 1543 & 18.910 & -0.539 & 0.029 & 18.876     & 0.136	 & 0.005    &  WHT2009sep21	   &	 DA		& 22.6/8.0 \\
44  & UVEXJ041745.78+454049.8  &  156.18824 &  -3.44853   & 1629 & 518 & 17.322 & 0.193 & 0.535 & 16.886       & 0.443	 & 0.172    &  WHT2010dec12	   &	 MS/BHB        & 13.4/3.9/F0V,0.0 \\
45  & UVEXJ041824.24+441152.2  &  157.30940 &  -4.42838   & 1635 & 519 & 18.132 & -0.112 & 0.599 & 17.640      & 0.472	 & 0.473    &  WHT2010dec13	   &	 CV		& 20/--20 \\
46  & UVEXJ041840.30+441714.1  &  157.28080 &  -4.33102   & 1635 & 1028 & 19.870 & -0.458 & 0.162 & 19.820     & -0.038  & -0.017   &  WHT2009sep23	   &	 DZA		&  \\
47  & UVEXJ041902.55+434307.1  &  157.72889 &  -4.68909   & 1628 & 1543 & 17.829 & -0.377 & -0.021 & 17.838    & 0.014	 & -0.123   &  WHT2010dec14	   &	 DA		& 17.4/8.0 \\
48  & UVEXJ041914.11+432147.8  &  158.00454 &  -4.91725   & 1628 & 1543 & 17.167 & -0.266 & 0.23 & 16.925      & 0.187	 & 0.017    &  WHT2010dec14	   &	 noisy  	& B8V,0.5 \\
49  & UVEXJ041926.84+440058.4  &  157.57106 &  -4.42618   & 1635 & 1543 & 19.073 & -0.611 & 0.13 & 18.986      & -0.007  & 0.158    &  WHT2010dec12	   &	 DC		& \\
50  & UVEXJ042023.52+473534.8  &  155.16783 &  -1.76858   & 1659 & 1543 & 16.055 & -0.998 & -0.057 & 16.204    & -0.013  & 0.128    &  WHT2009sep25/F	   &	 DAB		& \\
51  & UVEXJ042110.67+440945.6  &  157.68835 &  -4.10319   & 1661 & 1028 & 19.035 & -0.39 & 0.099 & 18.848      & 0.041	 & -0.163   &  WHT2009sep23	   &	 DA		& 20.5/7.8 \\
52  & UVEXJ042125.70+465115.4  &  155.81393 &  -2.16822   & 1672 & 514 & 18.107 & 0.48 & 0.877 & 17.451        & 0.637	 & 0.287    &  WHT2009sep21	   &	 sdB+F  	&  \\
53  & UVEXJ042223.30+440945.3  &  157.84196 &  -3.94941   & 1661 & 1543 & 19.397 & -0.443 & 0.239 & 19.134     & 0.240	 & -0.052   &  WHT2009sep24	   &	 noisy  	&  \\
54  & UVEXJ052823.37+275159.7  &  178.86901 &  -3.77121   & 2465 & 519 & 19.680 & -0.362 & 0.313 & 0.000       & no data & no data  &  WHT2010dec13	   &	 noisy  	& B3V,0.4 \\
55  & UVEXJ052825.82+320859.5  &  175.30149 &  -1.39485   & 2458 & 1543 & 18.501 & -0.32 & 0.093 & 0.000       & no data & no data  &  WHT2010dec12	   &	 DA		& 9.9/8.3 \\
56  & UVEXJ052835.30+271650.0  &  179.38341 &  -4.05755   & 2454 & 515 & 14.499 & 0.155 & 0.5 & 0.000	       & no data & no data  &  WHT2010dec12	   &	 BHB/MS        & 17.3/3.7/B3V,0.5 \\
57  & UVEXJ052847.75+322330.3  &  175.14281 &  -1.19658   & 2458 & 1028 & 18.876 & -0.06 & 0.082 & 0.000       & no data & no data  &  WHT2010dec13	   &	 DA		& 12.9/7.9 \\
58  & UVEXJ052851.01+262946.3  &  180.07233 &  -4.44047   & 2467 & 519 & 18.980 & 0.007 & 0.578 & 0.000        & no data & no data  &  WHT2010dec13	   &	 noisy       & B3V,0.6 \\
59  & UVEXJ185740.07+075557.3  &   40.70227 &  +2.23345   & 4346 & 1028 & 18.934 & -0.312 & 0.168 & 18.817     & 0.307	 & 0.058    &  Hect2006may02	   &	 DA		&  \\
60  & UVEXJ190812.07+164029.2  &   49.66841 &  +3.92988   & 4556 & 1028 & 17.292 & -0.619 & 0.051 & 17.157     & -0.102  & -0.074   &  WHT2010dec14	   &	 DA		& 17.0 \\
61  & UVEXJ190912.34+021342.8  &   36.94523 &  -2.93167   & 4580 & 519 & 18.764 & -0.218 & 0.867 & 18.220      & 1.290	 & 0.643    &  WHT2010dec13	   &	 DA+dM        & DA+M2Ve \\
62  & UVEXJ193809.18+305401.5  &   65.47669 &  +4.56471   & 5150 & 519 & 17.038 & -0.297 & 0.197 & 0.000       & 0.015	 & 0.112    &  WHT2009sep23	   &	 BHB/MS        & 18.9/4.5/B3V,0.4 \\
63  & UVEXJ193813.83+313708.1  &   66.11524 &  +4.89870   & 5129 & 1543 & 17.373 & -0.996 & -0.074 & 0.000     & -0.101  & 0.035    &  WHT2009sep23	   &	 sdO		& 50.4/5.7/--1.47 \\
64  & UVEXJ193847.06+312024.2  &   65.92808 &  +4.66012   & 5160 & 519 & 19.481 & -0.743 & 0.161 & 19.467      & 0.104	 & 0.088    &  WHT2009sep25	   &	 G0V-G2V	& G2V,0.3 \\
65  & UVEXJ193951.69+302600.7  &   65.24726 &  +4.01643   & 5182 & 519 & 17.358 & -0.502 & 0.199 & 17.201      & 0.118	 & 0.131    &  WHT2009sep24	   &	 sdB	    & 27.8/4.6/--1.5 \\
66  & UVEXJ194028.01+322039.8  &   66.98290 &  +4.83719   & 5186 & 1543 & 18.638 & -0.349 & 0.179 & 18.442     & 0.084	 & 0.236    &  WHT2009sep24	   &	 BHB/MS       & 16.0/3.1/B0V,0.4 \\
67  & UVEXJ194059.93+322347.8  &   67.08369 &  +4.76468   & 5186 & 515 & 19.237 & -0.725 & 0.67 & 18.687       & 0.820	 & 0.439    &  WHT2010dec13	   &	 DAB+dM        & DAB+M1V \\
68  & UVEXJ194135.69+321222.0  &   66.97892 &  +4.56159   & 5186 & 1028 & 18.892 & -0.82 & 0.065 & 18.891      & 0.190	 & 0.144    &  WHT2009sep23	   &	 sdB	    & 31.0/6.0/--1.9 \\
69  & UVEXJ194633.12+193926.4  &   56.64074 &  -2.64941   & 5284 &  519 & 15.135 & -0.461 & 0.331 & 14.809     & 0.282	 & 0.355    &  Fast2011jun09	   &	 CV	   &   \\
70  & UVEXJ202249.99+412423.1  &   79.15873 &  +2.42925   & 5853 & 1028 & 19.718 & -0.248 & 0.049 & 19.544     & -0.154  & -0.520   &  WHT2009sep24	   &	 DA		& 15.6/8.2 \\
71  & UVEXJ202255.55+412504.9  &   79.17820 &  +2.42162   & 5853 & 1028 & 18.766 & -0.307 & -0.003 & 18.651    & -0.080  & -0.259   &  WHT2009sep24	   &	 DA		& 16.3/8.0 \\
72  & UVEXJ202350.92+423826.0  &   80.28021 &  +2.98039   & 5875 & 1543 & 18.641 & -0.824 & -0.133 & 18.781    & -0.178  & 0.173    &  WHT2010dec12	   &	 DA		& 25.3/8.5 \\
73  & UVEXJ202439.91+400630.7  &   78.29391 &  +1.40076   & 5892 & 1031 & 19.905 & -0.749 & -0.178 & 20.005    & -0.252  & -0.080   &  WHT2009sep24	   &	 DA		& 24.0   \\
74  & UVEXJ202457.34+410804.1  &   79.16473 &  +1.94658   & 5879 & 1028 & 19.768 & -0.433 & -0.071 & 19.824    & -0.117  & -0.030   &  WHT2009sep24	   &	 DA		& 22.0   \\
75  & UVEXJ202501.86+411626.0  &   79.28692 &  +2.01532   & 5916 & 1028 & 18.627 & -0.305 & -0.005 & 18.703    & -0.113  & -0.166   &  WHT2009sep25	   &	 DA		& 14.8/8.2 \\
\hline   
\end{tabular}                                                                                
\end{sideways}
\end{table*}

\begin{table*}
\caption[]{Table \ \ref{tab:spectra} continued \label{tab:spectra2} }
\centering
\begin{sideways}
\tiny
\noindent
\begin{tabular}{p{3mm}@{\ \ \ }p{30mm}@{\ \ \ }p{10mm}@{\ \ \ }p{10mm}@{\ \ \ }p{6mm}@{\ \ \ }p{8mm}@{\ \ \ }p{7mm}@{\ \ \ }
p{7mm}@{\ \ \ }p{7mm}@{\ \ \ }p{7mm}@{\ \ \ }p{8mm}@{\ \ \ }p{10mm}@{\ \ \ }p{17mm}@{\ \ \ }p{16mm}@{\ \ \ }p{16mm}}
\hline 
No & Name & l & b & Field & Selection & $g$ & ($U-g$) & ($g-r$) & ($HeI$) & ($r-i$) & ($r-H\alpha$) & Run & Classification & Fit \\ \hline
76  & UVEXJ202557.21+400949.2  &   78.48112 &  +1.23151   & 5892 & 1028 & 18.429 & -0.735 & -0.204 & 18.469    & -0.164  &  -0.522   &  WHT2009sep24	    &	  DA		 & 30.0   \\
77  & UVEXJ202630.19+405024.1 &   79.09318 &  +1.53808    & 5902 & 1028 & 19.880 & -0.182 & 0.145 & 19.864     & -0.020  &  0.378    &  Hect2005oct23	   &	unknown/Em	   &  \\
78  & UVEXJ202659.21+411644.1 &   79.50370 &  +1.71848    & 5916 & 1028 & 17.092 & -0.703 & -0.083 & 17.077    & 0.073	 &  0.047    &  Hect2005oct22	   &	 DA		   &  \\
79  & UVEXJ202712.06+424720.1 &   80.75706 &  +2.56286    & 5947 & 519 & 16.079 & -0.385 & 0.33 & 15.864       & 0.176	 &  0.464    &  WHT2009sep21	    &	  CV(novalike)     & 24/--21 \\
80  & UVEXJ202744.63+405044.5 &   79.23419 &  +1.35043    & 5939 & 1543 & 19.134 & -0.814 & -0.084 & 19.363    & -0.152  &  0.041    &  Hect2005oct22	   &	 DB		   &  \\
81  & UVEXJ202800.47+405620.0 &   79.33903 &  +1.36420    & 5939 & 1028 & 16.167 & -0.754 & -0.157 & 16.258    & -0.147  &  -0.148   &  WHT2009sep21	    &	  DA		   & 20.0 \\
82  & UVEXJ202807.55+411357.7 &   79.59074 &  +1.51750    & 5951 & 1543 & 18.939 & -0.662 & -0.082 & 19.056    & 0.045	 &  -0.159   &  Hect2005oct22	   &	 DA		   &  \\
83  & UVEXJ202922.26+412815.9 &   79.92088 &  +1.46732    & 5951 & 1029 & 20.161 & -0.309 & 0.081 & 19.946     & -0.035  &  -0.017   &  Hect2005oct22	   &	 DA		   &  \\
84  & UVEXJ203039.82+413252.3 &   80.12526 &  +1.31662    & 5985 & 1543 & 19.436 & -0.373 & -0.04 & 19.386     & -0.104  &  -0.105   &  Hect2004jun10	   &	 DA		   &  \\
85  & UVEXJ203238.52+411339.4 &   80.08644 &  +0.82801    & 6010 & 1543 & 19.295 & -0.775 & -0.151 & 19.486    & -0.101  &  -0.127   &  Hect2004jun10	   &	 DA		   &  \\
86  & UVEXJ203352.73+405647.0 &   79.99913 &  +0.47336    & 6010 & 1028 & 18.461 & -0.978 & -0.199 & 18.573    & -0.202  &  0.031    &  Hect2006oct10	   &	 DA		   &  \\
87  & UVEXJ203411.72+411020.3 &   80.21601 &  +0.56034    & 6010 & 1543 & 20.428 & -0.5 & -0.015 & 20.333      & 0.212	 &  -0.395   &  Hect2006oct10	   &	 DA		   &  \\
88  & UVEXJ203413.57+404702.9 &   79.90816 &  +0.32382    & 6035 & 1029 & 19.615 & -0.309 & 0.034 & 19.681     & -0.095  &  -0.214   &  Hect2005jul01	   &	 DA		   &  \\
89  & UVEXJ203421.28+404827.4 &   79.94151 &  +0.31834    & 6035 & 1028 & 20.521 & -0.288 & 0.157 & 20.380     & -0.084  &  -0.182   &  Hect2005jul02	   &	 DA		   &  \\
90  & UVEXJ203455.77+412735.1 &   80.52884 &  +0.62177    & 6046 & 1028 & 20.884 & -0.457 & 0.71 & 0.000       & 0.109	 &  -0.131   &  Hect2006oct10	   &	 DA+dM     &  \\
91  & UVEXJ203519.00+403408.9 &   79.85970 &  +0.02950    & 6035 & 1029 & 19.580 & -0.511 & 0.001 & 19.587     & -0.103  &  -0.244   &  Hect2006oct10	   &	 DA		   &  \\
92  & UVEXJ203614.30+392309.8 &   79.02005 &  -0.82249    & 6036 & 1029 & 19.776 & -0.584 & 0.194 & 19.635     & 0.821	 &  0.326    &  WHT2009sep25	    &	  DA+dM 	   & DA+M3V\\
93  & UVEXJ203656.54+392934.9 &   79.18734 &  -0.86650    & 6066 & 1543 & 18.882 & -1.144 & -0.178 & 19.100    & -0.219  &  0.025    &  WHT2009sep23	    &	  noisy/WD?		   &  \\
94  & UVEXJ203739.63+413216.3 &   80.89908 &  +0.26049    & 6080 & 1028 & 19.645 & -0.376 & -0.002 & 19.779    & 0.162	 &  -0.198   &  WHT2009sep23/H      &	  DA		   & 17.0 \\
95  & UVEXJ204101.05+414327.9 &   81.42907 &  -0.12343    & 6112 & 1543 & 19.682 & -0.299 & 0.289 & 19.410     & 0.154	 &  0.109    &  Hect2004jun19	   &	 DA		   &  \\
96  & UVEXJ204210.22+443928.7 &   83.87361 &  +1.51222    & 6107 & 519 & 18.738 & 0.002 & 0.891 & 18.055       & 0.712	 &  0.513    &  WHT2010dec13	    &	  sdO?  	   & B0V,0.9 \\
97  & UVEXJ204649.51+410906.2 &   81.65384 &  -1.33263    & 6186 & 1543 & 20.158 & -0.323 & 0.178 & 20.186     & 0.165	 &  0.239    &  Hect2004jun25	   &	 DZ		   &  \\
98  & UVEXJ204710.61+413133.5 &   81.98697 &  -1.14960    & 6166 & 1543 & 18.945 & -1.032 & -0.137 & 19.311    & -0.115  &  -0.031   &  Hect2004jun26	   &	 DA		   &  \\
99  & UVEXJ204751.27+442920.1 &   84.37091 &  +0.61442    & 6196 & 1028 & 19.960 & -0.503 & -0.044 & 20.165    & 0.242	 &  0.020    &  Hect2004nov20	   &	 DA		   &  \\
100 & UVEXJ204758.19+382323.2 &   79.63810 &  -3.23534    & 6181 & 519 & 19.054 & -0.239 & 0.639 & 18.605      & 0.454	 &  0.267    &  Hect2004nov20	   &	 noisy     &  \\
101 & UVEXJ204923.48+381139.0 &   79.66150 &  -3.57521    & 6211 & 1542 & 18.891 & -0.126 & 0.497 & 18.512     & 0.266	 &  0.463    &  Hect2004nov12	   &	 unknown/Em		   &  \\
102 & UVEXJ204925.78+411725.6 &   82.06990 &  -1.62608    & 6217 & 1543 & 20.809 & 0.06 & 0.573 & 20.196       & 0.533	 &  0.189    &  Hect2004jun25	   &	 DC		   &  \\
103 & UVEXJ204932.62+374048.2 &   79.28056 &  -3.92266    & 6201 & 1028 & 18.577 & -0.535 & -0.027 & 18.519    & 0.031	 &  -0.194   &  Hect2004nov12	   &	 DA		   &  \\
104 & UVEXJ204945.83+382057.2 &   79.82825 &  -3.53410    & 6211 & 1543 & 19.749 & -0.374 & 0.035 & 19.527     & -0.371  &  0.297    &  Hect2004nov12	   &	 DA		   &  \\
105 & UVEXJ204957.75+401637.9 &   81.34841 &  -2.34506    & 6226 & 519 & 15.741 & -0.67 & 0.334 & 15.485       & 0.206	 &  0.179    &  WHT2010dec13	    &	  He-sdO	   & 45.9/6.1/1.4 \\
106 & UVEXJ205037.81+424618.9 &   83.35761 &  -0.85988    & 6220 & 1543 & 15.758 & -0.571 & -0.042 & 15.730    & -0.072  &  -0.144   &  WHT2008oct05	   &	 DA		   & 17.0 \\
107 & UVEXJ205039.07+373958.4 &   79.40858 &  -4.10129    & 6218 & 1031 & 21.119 & -0.535 & 0.283 & 18.981     & 0.098	 &  2.248    &  Hect2004nov12	   &	 M5III  	   & M5III \\
108 & UVEXJ205056.87+380710.6 &   79.79745 &  -3.85943    & 6231 & 519 & 20.404 & -0.285 & 0.588 & 20.150      & 0.369	 &  0.355    &  Hect2004nov20	   &	 DAB/DA 	   &  \\
109 & UVEXJ205148.13+442408.8 &   84.75023 &  +0.01441    & 6219 & 1028 & 17.716 & -0.888 & -0.098 & 17.874    & -0.120  &  0.015    &  WHT2009sep22	    &	  DA		   & 26.0 \\
110 & UVEXJ205449.65+371953.2 &   79.67935 &  -4.95297    & 6267 & 519 & 18.396 & -0.471 & 0.606 & 17.971      & 1.848	 &  0.956    &  WHT2010dec13	    &	  TTauri(M5V)	   & 4/--9 \\
111 & UVEXJ210037.77+501029.0 &   90.11721 &  +2.61574    & 6323 & 1543 & 16.542 & -0.735 & -0.154 & 16.626    & -0.283  &  -0.092   &  WHT2010dec14	    &	  DA		   & 20.9/7.9 \\
112 & UVEXJ210127.26+452247.2 &   86.60349 &  -0.65097    & 6355 & 519 & 18.185 & 0.044 & 0.597 & 0.000        & 0.571	 &  0.562    &  WHT2009sep21	    &	  CV		   & 28/--22 \\
113 & UVEXJ210248.44+475058.9 &   88.60784 &  +0.81096    & 6341 & 1543 & 18.300 & -0.421 & -0.013 & 18.176    & no data &   no data &  WHT2010dec12	    &	  DA		   & 13.3/8.1 \\
114 & UVEXJ210454.41+460041.9 &   87.47634 &  -0.68123    & 6365 & 1543 & 18.743 & -0.837 & -0.037 & 18.914    & -0.177  &  0.221    &  WHT2010dec12	    &	  DB		   & 16.0 \\
115 & UVEXJ211718.18+550638.7 &   95.46871 &  +4.09795    & 6496 & 1543 & 17.342 & -0.597 & -0.078 & 17.335    & -0.062  &  -0.087   &  WHT2009sep22	    &	  DA		   & 22.0 \\
116 & UVEXJ212257.82+552609.0 &   96.26868 &  +3.75358    & 6555 & 519 & 15.057 & -0.13 & 0.455 & 14.822       & 1.318	 &    0.512  &  WHT2009sep22	    &	  DA+dM 	   & DA+M1V\\
117 & UVEXJ212409.05+555521.4 &   96.73083 &  +3.98237    & 6561 & 1028 & 17.952 & -0.245 & 0.046 & 17.719     & no data &   no data &  WHT2009sep21	    &	  DA		   & 12.0 \\
118 & UVEXJ212705.33+541058.2 &   95.81804 &  +2.44184    & 6597 & 1543 & 18.929 & -0.799 & -0.042 & 19.150    & -0.117  &  0.147    &  WHT2010dec13	    &	  DB		   & 18.0 \\
119 & UVEXJ212852.14+542048.4 &   96.11948 &  +2.38077    & 6603 & 1028 & 18.126 & -0.395 & -0.047 & 18.030    & 0.003	 &  -0.162   &  WHT2010dec13	    &	  DA		   & 13.9/8.2 \\
120 & UVEXJ222940.17+610700.7  &  106.64583 &  +2.81053   & 7084 & 1543 & 17.489 & -0.789 & -0.111 & 17.553    & 0.001	 &  -0.022   &  WHT2010dec14	    &	  DA		   & 22.3/8.0 \\
121 & UVEXJ223634.77+591907.8  &  106.47749 &  +0.82554   & 7139 & 1028 & 17.647 & -0.724 & -0.081 & 17.645    & -0.018  &  -0.023   &  WHT2010dec14	    &	  DA		   & 28.7/8.0 \\
122 & UVEXJ223811.54+603759.9  &  107.29952 &  +1.87109   & 7155 & 1543 & 18.533 & -0.65 & 0.14 & 18.350       & 0.003	 &  -0.166   &  WHT2010dec13	    &	  DA		   & 12.6/8.1 \\
123 & UVEXJ223941.98+585729.1  &  106.65014 &  +0.31521   & 7153 & 515 & 16.020 & -0.289 & 0.496 & 15.594      & 0.297	 &  0.228    &  WHT2010dec13	    &	  He-sdO	   & 47.7/5.0/--0.74 \\
124 & UVEXJ224010.23+555950.6  &  105.27455 &  -2.30746   & 7171 & 1028 & 17.741 & -0.651 & -0.086 & 17.796    & -0.028  &  -0.061   &  WHT2010dec10	    &	  DA		   & 20.0/3.0 \\
125 & UVEXJ224112.21+564419.1  &  105.75676 &  -1.72636   & 7179 & 518 & 18.978 & -0.078 & 0.56 & 18.634       & 0.428	 &  0.545    &  Hect2005jul05	   &	 CV(magnetic)	   &  \\
126 & UVEXJ224145.94+562230.0  &  105.65079 &  -2.08275   & 7171 & 514 & 17.981 & -0.221 & 0.561 & 17.569      & 0.385	 &  0.342    &  WHT2010dec13	    &	  DAe	   &  \\
127 & UVEXJ224324.38+573324.0  &  106.40874 &  -1.14878   & 7173 & 1543 & 19.666 & -0.492 & 0.381 & 19.274     & 0.215	 &  0.006    &  WHT2010dec13	    &	  noisy    &  \\
128 & UVEXJ224435.05+550104.9  &  105.35839 &  -3.46612   & 7188 & 514 & 14.917 & -0.039 & 0.388 & 14.618      & 0.308	 &  0.495    &  Fast2009nov21	   &	 Be		   &  \\
129 & UVEXJ224521.39+551705.3  &  105.58100 &  -3.28184   & 7197 & 1028 & 19.632 & -0.661 & -0.033 & 19.605    & -0.176  &  -0.168   &  WHT2010dec14	    &	  MS/BHB	   & 14.9/4.7/F0V,0.1 \\
130 & UVEXJ224610.82+611450.3  &  108.44539 &  +1.94933   & 7189 & 1028 & 17.974 & -0.456 & -0.088 & 18.009    & -0.005  &  -0.059   &  WHT2010dec14	    &	  DA		   & 20.0/8.1 \\
131 & UVEXJ224721.36+565937.4  &  106.61932 &  -1.89546   & 7203 & 518 & 17.736 & -0.114 & 0.468 & 17.356      & 0.399	 &  0.257    &  WHT2010dec14	    &	  DAe	   &  \\
132 & UVEXJ224835.69+605057.8  &  108.52252 &  +1.46171   & 7206 & 1028 & 17.446 & -0.779 & 0.047 & 17.375     & 0.008	 &  0.066    &  WHT2010dec10	    &	  noisy 	   & O5V,0.4 \\
\hline   
\end{tabular}                                                                                
\end{sideways}
\end{table*}

\newpage 

\begin{figure*}
\centerline{\epsfig{file=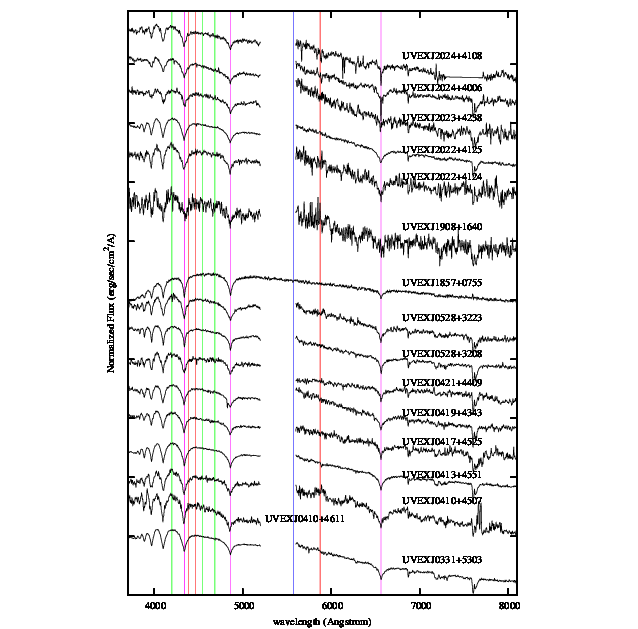,width=22cm,angle=0,clip=}}
\caption{DA white dwarfs. The pink, red and green lines indicate the position of hydrogen, HeI and HeII respectively. The blue line is a skyline in the
Hectospec spectra.
\label{fig:spectra2}}
\end{figure*}

\begin{figure*}
\centerline{\epsfig{file=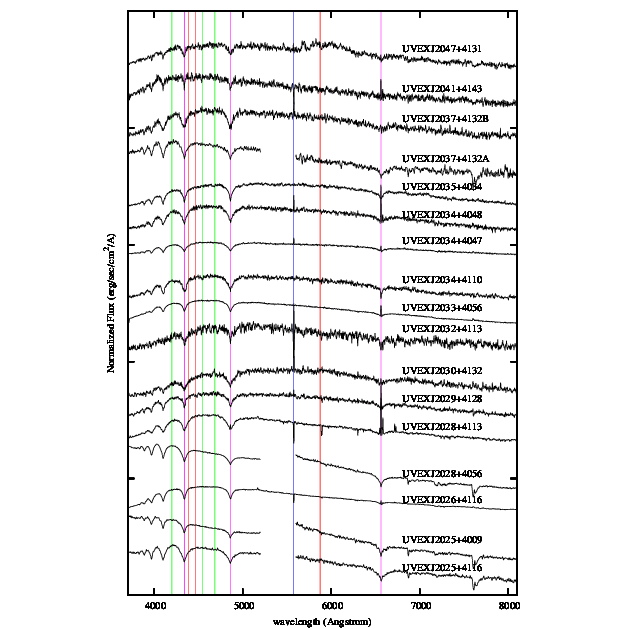,width=22cm,angle=0,clip=}}
\caption{DA white dwarfs. The pink, red and green lines indicate the position of hydrogen, HeI and HeII respectively. The blue line is a skyline in the
Hectospec spectra.
\label{fig:spectra3}}
\end{figure*}

\begin{figure*}
\centerline{\epsfig{file=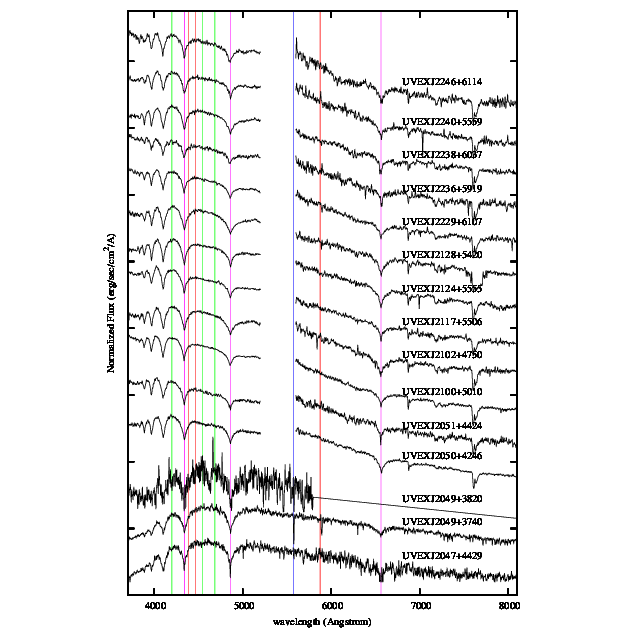,width=22cm,angle=0,clip=}}
\caption{DA white dwarfs. The pink, red and green lines indicate the position of hydrogen, HeI and HeII respectively. The blue line is a skyline in the
Hectospec spectra.
\label{fig:spectra4}}
\end{figure*}

\begin{figure*}
\centerline{\epsfig{file=AppendixPlot5DBWDs.eps,width=22cm,angle=0,clip=}}
\caption{All UV-excess spectra classified as DB and DAB white dwarfs. The pink, red and green lines indicate the position of hydrogen, 
HeI and HeII respectively. The blue line is a skyline in the Hectospec spectra.
\label{fig:spectra5}}
\end{figure*}

\begin{figure*}
\centerline{\epsfig{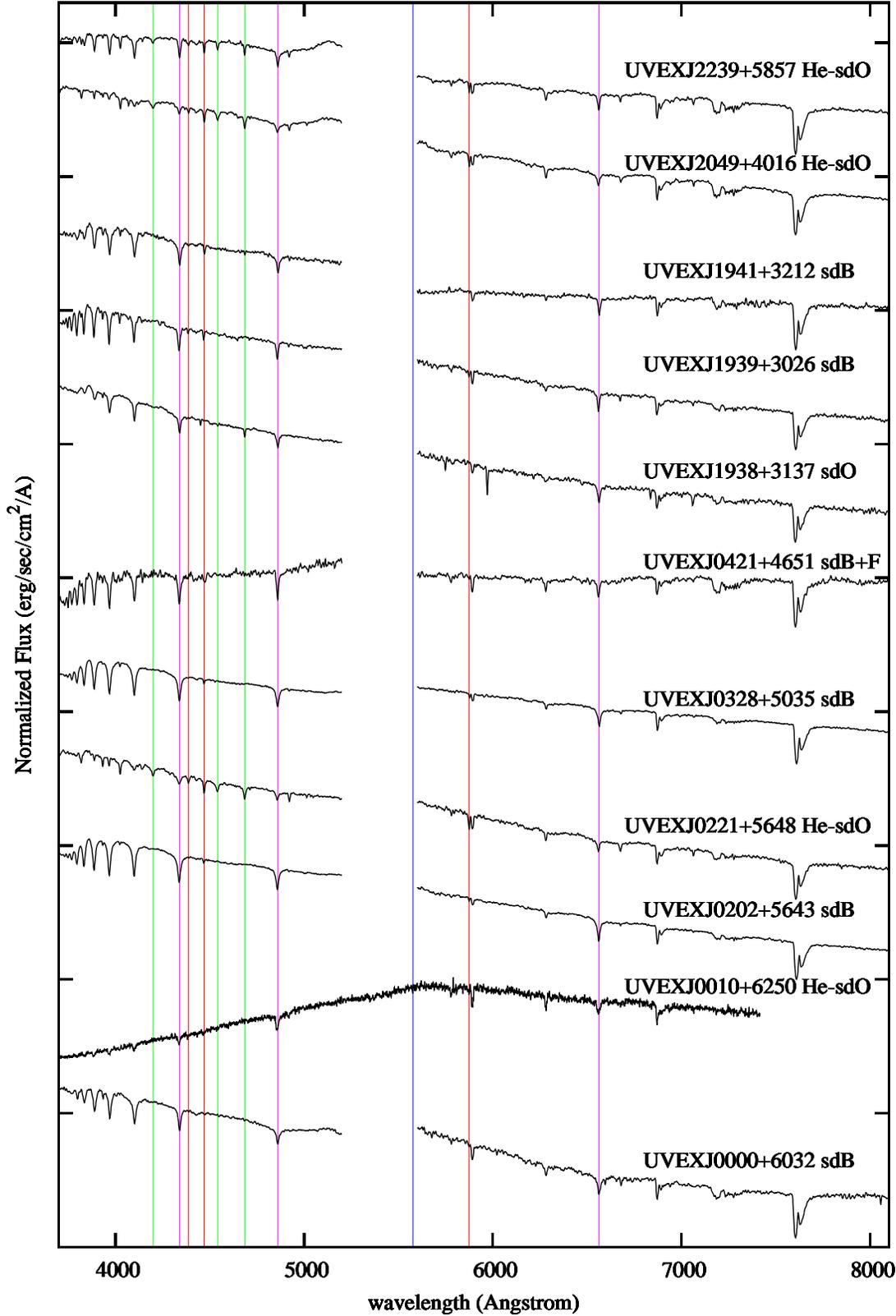}}
\caption{The UV-excess spectra classified as He-sdO, sdO, sdB and sdB+F candidates. The pink, red and green lines indicate the position of hydrogen, 
HeI and HeII respectively. The blue line is a skyline in the Hectospec spectra.
\label{fig:spectra6}}
\end{figure*}

\begin{figure*}
\centerline{\epsfig{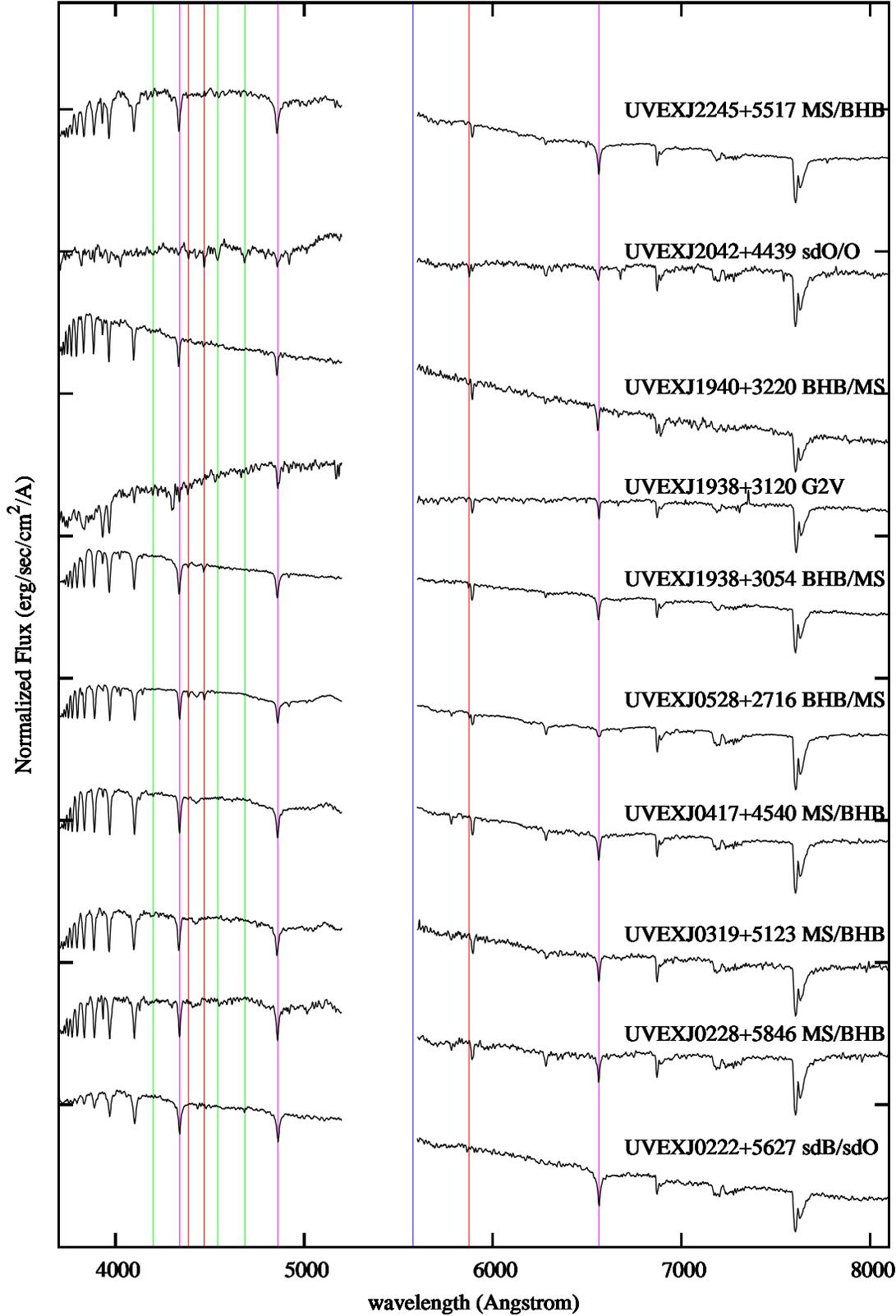}}
\caption{The UV-excess spectra classified as MS/BHB stars, probably sdB/sdO/O type stars and 1 G-type star. 
The pink, red and green lines indicate the position of hydrogen, HeI and HeII respectively. The blue line is a skyline in the Hectospec spectra.
\label{fig:spectra7}}
\end{figure*}

\begin{figure*}
\centerline{\epsfig{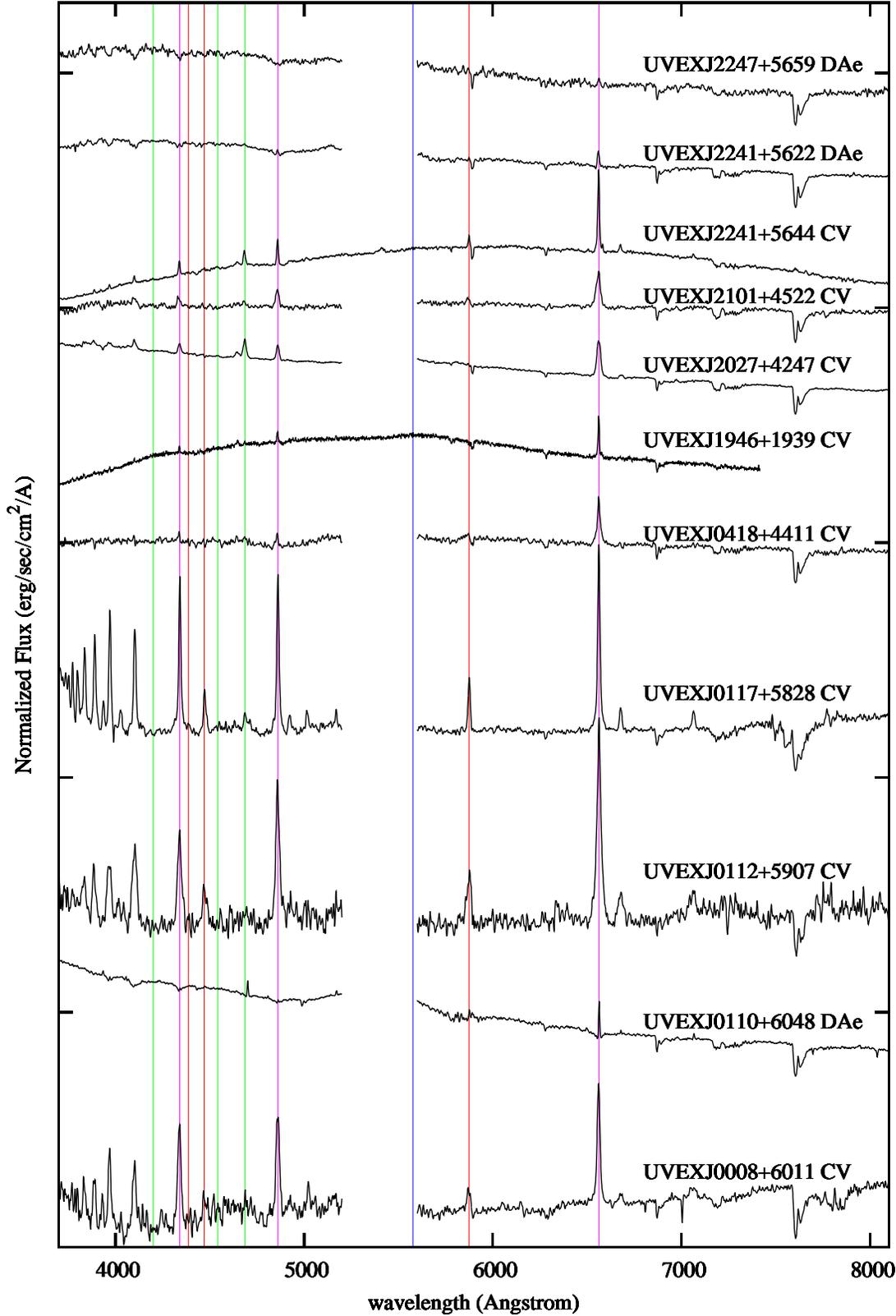}}
\caption{All UV-excess spectra classified as Cataclysmic Variables and DAe white dwarfs. 
The pink, red and green lines indicate the position of hydrogen, HeI and HeII respectively. The blue line is a skyline in the Hectospec spectra.
\label{fig:spectra8}}
\end{figure*}

\begin{figure*}
\centerline{\epsfig{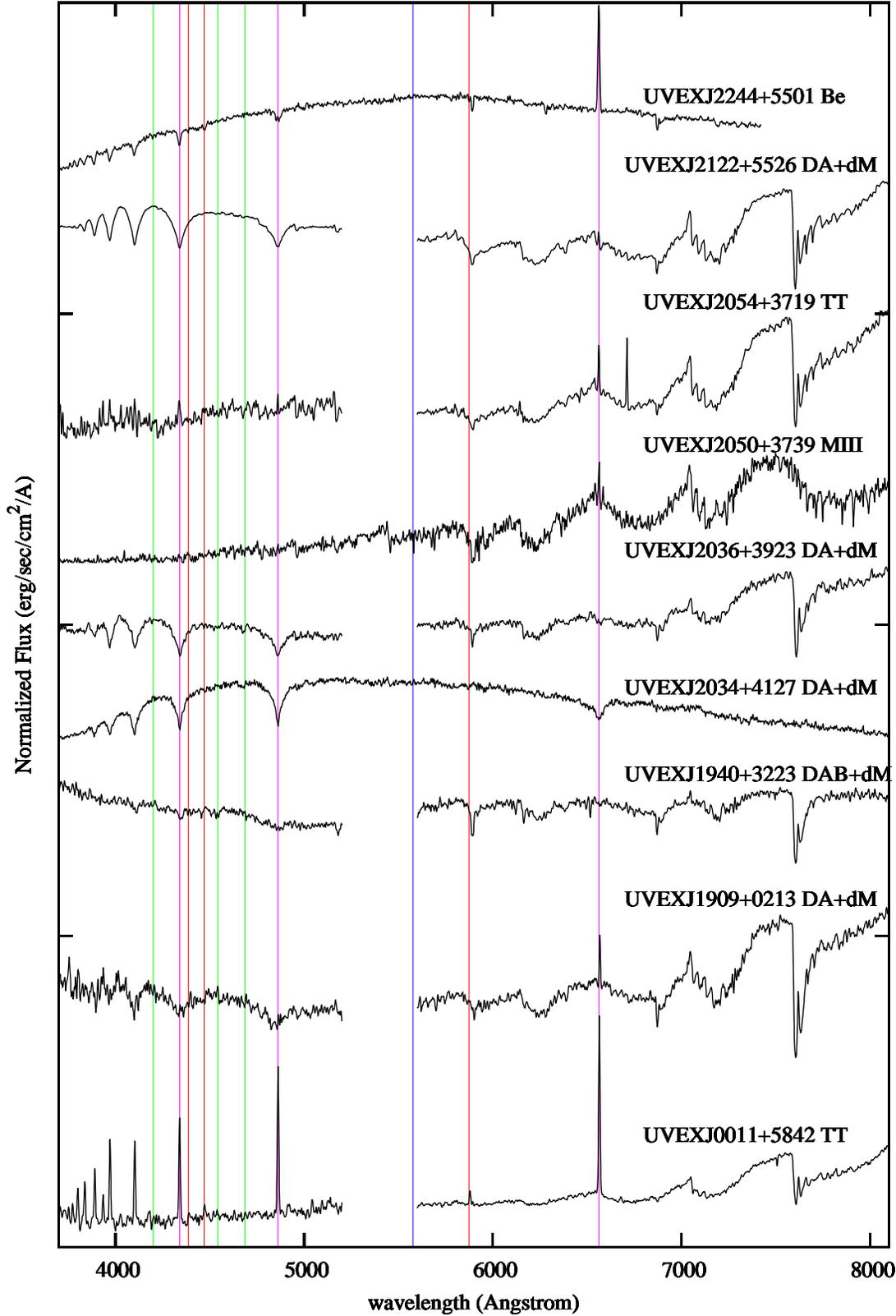}}
\caption{The UV-excess spectra classified as DA+dM systems, T Tauri stars, Be star and 1 M5III giant.
The pink, red and green lines indicate the position of hydrogen, HeI and HeII respectively. The blue line is a skyline in the Hectospec spectra.
\label{fig:spectra9}}
\end{figure*}

\begin{figure*}
\centerline{\epsfig{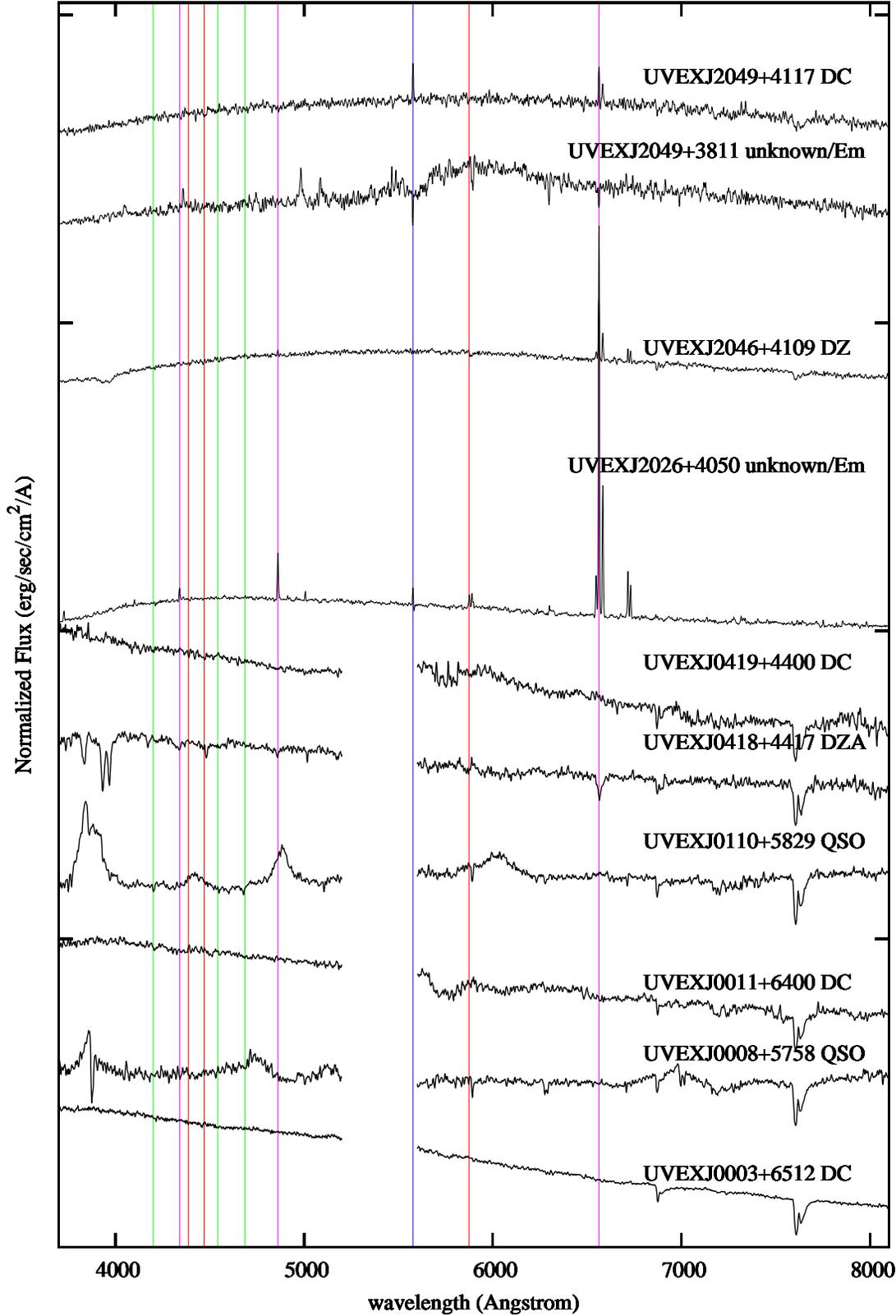}}
\caption{UV-excess spectra classified as DC and DZ white dwarfs, QSOs and 2 unknown sources. Note that the emission lines of the 2 unknown sources, also present in the
sky offset spectra, are due to diffuse emission in the field.
The pink, red and green lines indicate the position of hydrogen, HeI and HeII respectively. The blue line is a skyline in the Hectospec spectra.
\label{fig:spectra10}}
\end{figure*}

\begin{figure*}
\centerline{\epsfig{file=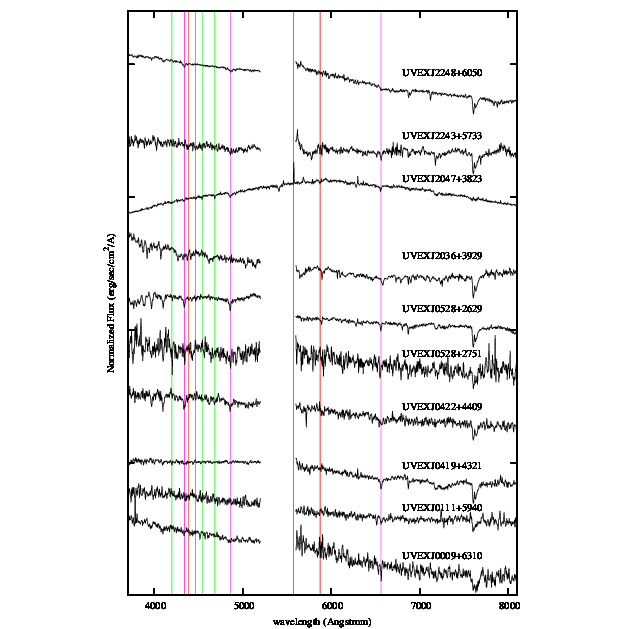,width=22cm,angle=0,clip=}}
\caption{UV-excess spectra classified as ``Noisy''.
The pink, red and green lines indicate the position of hydrogen, HeI and HeII respectively. The blue line is a skyline in the Hectospec spectra.
\label{fig:spectra11}}
\end{figure*}

\label{lastpage}


\newpage

\begin{thebibliography}{99}

\bibitem[\protect\citeauthoryear{Aungwerojwit et al.}{2005}]{aungwerojwit05}
Aungwerojwit A., G\"ansicke B.T., Rodriguez-Gil P., et al., 2005, A\&A 443, 995A
\bibitem[\protect\citeauthoryear{Barentsen et al.}{2011}]{barentsen11}
Barentsen G., Vink J.S., Drew J.E., Greimel R. et al., 2011, MNRAS 415, 103B
\bibitem[\protect\citeauthoryear{Bergeron et al.}{1992}]{bergeron92}
Bergeron P., Saffer R.A., Liebert J., 1992, ApJ 394, 228B
\bibitem[\protect\citeauthoryear{Bergeron et al.}{1995}]{bergeron95}
Bergeron P., Wesemael F. \& Beauchamp A., 1995, PASP 107, 1047
\bibitem[\protect\citeauthoryear{Brunzendorf et al.}{2002}]{brunzendorf02}
Brunzendorf J. \& Meusinger H., 2002, A\&A 390, 879B
\bibitem[\protect\citeauthoryear{Cardelli et al.}{1989}]{ccm89}
Cardelli J.A., Clayton G.C. \& Mathis J.S., 1989, ApJ 345, 245
\bibitem[\protect\citeauthoryear{Christlieb et al.}{2001}]{christlieb01}
Christlieb N., Wisotzki L., Reimers D., et al., 2001, A\&A 366, 898C
\bibitem[\protect\citeauthoryear{Corradi et al.}{2010}]{corradi10}
Corradi R. L. M., Valentini M., Munari U., Drew J. E., et al., 2010, A\&A 509, 41
\bibitem[\protect\citeauthoryear{Cutri et al.}{2003}]{cutri03}
Cutri R. M., Skrutskie M. F., van Dyk S. et al., 2003, yCat 2246, 0
\bibitem[\protect\citeauthoryear{Deacon et al.}{2009}]{deacon09}
Deacon N. R., Groot P. J., Drew J. E., et al., 2009, MNRAS 397, 1685
\bibitem[\protect\citeauthoryear{Demers et al.}{1986}]{demers86}
Demers S., Beland S., Kibblewhite E.J., et al., 1986, AJ 92, 878D
\bibitem[\protect\citeauthoryear{Drew et al.}{2005}]{drew05}
Drew J., Greimel R., Irwin M., et al., 2005, MNRAS 362, 753 (D05)
\bibitem[\protect\citeauthoryear{Downes}{1986}]{downes86}
Downes R. A., 1986, ApJS 61, 569D
\bibitem[\protect\citeauthoryear{Eisenstein et al.}{2006}]{eisenstein06}
Eisenstein D. J., Liebert J., Harris H. C., et al., 2006, ApJ 167, 40E
\bibitem[\protect\citeauthoryear{Eracleous et al.}{2002}]{eracleous02}
Eracleous M., Wade R. A., Mateen M., et al., 2002, PASP 114, 207E
\bibitem[\protect\citeauthoryear{Fabricant et al.}{1998}]{fabricant98}
Fabricant D., Cheimets P., Caldwell N., et al., 1998, PASP 110,79F
\bibitem[\protect\citeauthoryear{Fabricant et al.}{2004}]{fabricant04}
Fabricant et al., 2004, SPIE 5492, 767F
\bibitem[\protect\citeauthoryear{Fabricant et al.}{2005}]{fabricant05}
Fabricant D., Fata R., Rollet J., et al., 2005
\bibitem[\protect\citeauthoryear{Finley et al.}{1997}]{finley97}
Finley D.S., Koester D. \& Basri G., 1997, ApJ 488, 375F
\bibitem[\protect\citeauthoryear{G\"ansicke et al.}{2009}]{gansicke09}
G\"ansicke B.T., Dillon M., Southworth J., et al., 2009, MNRAS 397, 2170G
\bibitem[\protect\citeauthoryear{Gianninas et al.}{2011}]{gianninas11}
Gianninas A., Bergeron P., Ruiz M. T., 2011, ApJ 743, 138G
\bibitem[\protect\citeauthoryear{Girven et al.}{2011}]{girven11}
Girven J., G\"ansicke B.T., Steeghs D. \& Koester D., 2011, MNRAS 417, 1210G
\bibitem[\protect\citeauthoryear{Gonz\'{a}lez-Solares et al.}{2008}]{gonzalez2008}
Gonz\'{a}lez-Solares E.A., Walton N.A., Greimel R., Drew, J.E., et al., 2008, MNRAS 388, 89 
\bibitem[\protect\citeauthoryear{Green et al.}{1986}]{green86}
Green R. F., Schmidt M., Liebert J., 1986, ApJS 61, 305G
\bibitem[\protect\citeauthoryear{Greiss et al.}{2012}]{greiss12}
Greiss S., Steeghs D., G\"ansicke B.T., Martín E.L., Groot P.J. et al., 2012, arXiv1202.6333G
\bibitem[\protect\citeauthoryear{Groot et al.}{2009}]{groot09}
Groot P.J., Verbeek K., Greimel R., et al., 2009, MNRAS 399, 323G
\bibitem[\protect\citeauthoryear{Hagen et al.}{1995}]{hagen95}
Hagen H.-J., Groote D., Engels D., Reimers, D., 1995, A\&AS 111,195H
\bibitem[\protect\citeauthoryear{Harris et al.}{2003}]{harris03}
Harris H.C., Liebert J., Kleinman S.J. et al., 2003, AJ 126, 1023H
\bibitem[\protect\citeauthoryear{Holberg et al.}{2008}]{holberg08}
Holberg J.B., Sion E.M., Oswalt T., McCook G.P. et al., 2008, AJ 135,1225H
\bibitem[\protect\citeauthoryear{Homeier et al.}{1998}]{homeier98}
Homeier D., Koester D., Hagen H.J., et al., 1998 A\&A 338, 563H
\bibitem[\protect\citeauthoryear{Im et al.}{2007}]{im07}
Im Myungshin, Lee Induk., Yunseok C., et al. 2007, ApJ 664,64
\bibitem[\protect\citeauthoryear{Kepler et al.}{2007}]{kepler07}
Kepler S.O., Kleinman S.J., Nitta A., Koester D. et al., 2007, MNRAS 375, 1315K
\bibitem[\protect\citeauthoryear{Kilkenny et al.}{1997}]{kilkenny97}
Kilkenny D., O'Donoghue D, Koen C., et al., 1997, MNRAS 287, 867K
\bibitem[\protect\citeauthoryear{Knigge et al.}{2008}]{knigge08}
Knigge C, Scaringi S, Goad M.R., Cottis C.E., 2008, MNRAS 386, 1426K
\bibitem[\protect\citeauthoryear{Koester et al.}{2001}]{koester01}
Koester D., et al., 2001, A\&A, 378, 556
\bibitem[\protect\citeauthoryear{Krzesinski et al.}{2004}]{krzesinski04}
Krzesinski J., Nitta A., Kleinman S.J. et al., 2004, A\&A 417, 1093K
\bibitem[\protect\citeauthoryear{Lamontagne et al.}{2000}]{lamontagne00}
Lamontagne R., Demers S., Wesemael F., et al., 2000, AJ 119, 241L
\bibitem[\protect\citeauthoryear{Lanning}{1973}]{lanning73}
Lanning H.H., 1973, PASP 85, 70L
\bibitem[\protect\citeauthoryear{Lanning}{1982}]{lanning82}
Lanning H. H., 1982, ApJ 253,752L
\bibitem[\protect\citeauthoryear{Lanning}{2004}]{lanning04}
Lanning H. H., Meakes, M., 2004, PASP 116,1039L
\bibitem[\protect\citeauthoryear{Lee et al.}{2008}]{lee08}
Lee Induk, Im Myungshin, Kim Minjin et al., 2008, ApJS 175, 116L
\bibitem[\protect\citeauthoryear{L\'epine et al.}{2011}]{lepine11}
L\'epine S., Bergeron P., Lanning H. H., 2011, AJ 141, 96L
\bibitem[\protect\citeauthoryear{Liebert et al.}{2008}]{liebert05}
Liebert J., Bergeron P., Holberg J. B., 2005, ApJS 156, 47L
\bibitem[\protect\citeauthoryear{Limoges et al.}{2010}]{limoges00}
Limoges M., Bergeron P., 2010, ApJ 714, 1037L
\bibitem[\protect\citeauthoryear{McCook et al.}{1999}]{mcCook99}
McCook G. P., Sion E. M., 1999, ApJS 121, 1M
\bibitem[\protect\citeauthoryear{Miszalski et al.}{2008}]{miszalski08}
Miszalski B., Parker Q.A., Acker A. et al., 2008, MNRAS 384, 525M
\bibitem[\protect\citeauthoryear{Moe et al.}{2006}]{moe06}
Moe M., De Marco O., 2006, ApJ 650, 916
\bibitem[\protect\citeauthoryear{Moehler et al.}{1990}]{moehler90}
Moehler S., Richtler T., de Boer K.S. et al., 1990, A\&AS 86, 53M
\bibitem[\protect\citeauthoryear{Morales-Rueda \& Marsh}{2002}]{morales-rueda02}
Morales-Rueda L. \& Marsh T.R., 2002, MNRAS 332, 814M
\bibitem[\protect\citeauthoryear{Morgan et al.}{1943}]{morgan43}
Morgan W.W., Keenan P.C., Kellman E., 1943, QB881, M6
\bibitem[\protect\citeauthoryear{Napiwotzki}{1997}]{napiwotzki97}
Napiwotzki R., 1997, A\&A 322, 256N
\bibitem[\protect\citeauthoryear{Napiwotzki et al.}{1999}]{napiwotzki99}
Napiwotzki R., Green P.J., Saffer R.A., 1999, ApJ 517, 399N
\bibitem[\protect\citeauthoryear{Napiwotzki et al.}{2003}]{napiwotzki03}
Napiwotzki R., Christlieb N., Drechsel H., et al., 2003, Msngr 112, 25N
\bibitem[\protect\citeauthoryear{Nelemans et al.}{2001}]{nelemans01}
Nelemans G., Portegies Zwart S. F., Verbunt F., Yungelson L. R., 2001, A\&A 368, 939N
\bibitem[\protect\citeauthoryear{{\O}stensen et al.}{2011}]{ostensen11}
{\O}stensen R.H., Silvotti R., Charpinet S., et al., 2011, MNRAS 414, 2860O
\bibitem[\protect\citeauthoryear{Parker et al.}{2006}]{parker06}
Parker Q.A., Acker A., Frew D.J. et al., 2006, MNRAS 373, 79P
\bibitem[\protect\citeauthoryear{Pickles A.J.}{1998}]{pickles98}
Pickles A.J., 1998, PASP 110, 863 
\bibitem[\protect\citeauthoryear{Rafanelli}{1979}]{rafanelli79}
Rafanelli P., 1979, A\&A 76, 365R
\bibitem[\protect\citeauthoryear{Roeser et al.}{2010}]{roeser10}
Roeser S., Demleitner M. and Schilbach E., 2010, AJ 139,2440R
\bibitem[\protect\citeauthoryear{Sale et al.}{2009}]{sale09}
Sale S., Drew J., Unruh Y., et al., 2009, MNRAS 392, 497
\bibitem[\protect\citeauthoryear{Schlegel, Finkbeiner\& Davis}{1998}]{schlegel98}
Schlegel D.J., Finkbeiner D.P. \& Davis, M., 1998, ApJ 500, 525 
\bibitem[\protect\citeauthoryear{Silvestri et al.}{2006}]{Silvestri09}
Silvestri N M., Hawley S.L., West A.A., Szkody P. et al., 2006, AJ 131, 1674S
\bibitem[\protect\citeauthoryear{Sion et al.}{1983}]{sion83}
Sion E.M., Greenstein J.L., Landstreet J.D., Liebert J., et al., 1983, ApJ 269, 253S
\bibitem[\protect\citeauthoryear{Sion et al.}{1990}]{sion90}
Sion E.M., Kenyon S.J., Aannestad P.A., 1990, ApJS 72,707S
\bibitem[\protect\citeauthoryear{Spogli et al.}{1998}]{spogli98}
Spogli C., Fiorucci M. \& Tosti G., 1998, A\&AS 130, 485S
\bibitem[\protect\citeauthoryear{Stobie et al.}{1987}]{stobie87}
Stobie R.S., Morgan D.H., Bhatia R.K., et al., 1987, fbs conf, 493S
\bibitem[\protect\citeauthoryear{Stobie et al.}{1997}]{stobie97}
Stobie R.S., Kilkenny D., O'Donoghue D., et al., 1997, MNRAS 287, 848S
\bibitem[\protect\citeauthoryear{Szkody et al.}{2005}]{szkody05}
Szkody P., Henden A., Mannikko L., 2007, AJ 134, 185S
\bibitem[\protect\citeauthoryear{Tappert et al.}{2009}]{Tappert09}
Tappert C., G\"ansicke B.T., Zorotovic M., et al., 2009, A\&A 504, 491T
\bibitem[\protect\citeauthoryear{Verbeek et al.}{2011}]{verbeek12}
Verbeek K., Groot P.J., de Groot E., Scaringi S., Drew J.E., et al., 2012, MNRAS 420, 1115V
\bibitem[\protect\citeauthoryear{Vink et al.}{2008}]{vink08}
Vink J.S., Drew J.E., Steeghs D., Wright N.J., 2008, MNRAS 387, 308V
\bibitem[\protect\citeauthoryear{Wegner et al.}{1987}]{wegner87}
Wegner G., McMahan R.K., Boley F.I., 1987, AJ 94, 1271W
\bibitem[\protect\citeauthoryear{Wesemael et al.}{1993}]{wesemael93}
Wesemael F., Greenstein J.L., Liebert J., et al., 1993, PASP 105, 761W
\bibitem[\protect\citeauthoryear{Wisotzki et al.}{1996}]{wisotzki96}
Wisotzki L., Koehler T., Groote D. \& Reimers D., 1996, A\&AS 115, 227W
\bibitem[\protect\citeauthoryear{Witham et al.}{2008}]{witham08}
Witham A.R., Knigge C., Drew J.E., et al., 2008, MNRAS 384, 1277 
\bibitem[\protect\citeauthoryear{Yanny et al.}{2009}]{yanny09}
Yanny B., Rockosi C., Newberg H.J., et al., 2009, AJ 137, 4377
\bibitem[\protect\citeauthoryear{York et al.}{2000}]{york00}
York D.G., Adelman J., Anderson J.E. et al., 2000, AJ 120, 1579Y
\end{thebibliography}
\end{document}